\documentclass[longauth]{aa}  
\usepackage{ifpdf}
\usepackage{graphicx}
\usepackage{fixltx2e} 

\usepackage{txfonts}
\usepackage{natbib}                
\bibpunct{(}{)}{;}{a}{}{,}         
\usepackage{verbatim}

\begin{document}
\title{A multiwavelength view of the flaring state \\ of PKS~2155-304 in 2006}

   \titlerunning{A MWL view of the flaring state of PKS2155-304 in 2006}

\author{HESS Collaboration
\and A.~Abramowski \inst{1}
\and F.~Acero \inst{2}
\and F.~Aharonian \inst{3,4,5}
\and A.G.~Akhperjanian \inst{6,5}
\and G.~Anton \inst{7}
\and A.~Balzer \inst{7}
\and A.~Barnacka \inst{8,9}
\and U.~Barres~de~Almeida \inst{10}\thanks{supported by CAPES Foundation, Ministry of Education of Brazil}
\and Y.~Becherini \inst{11,12}
\and J.~Becker \inst{13}
% ...
\newpage
}

\offprints{Andreas Zech, Catherine Boisson}

\institute{
Universit\"at Hamburg, Institut f\"ur Experimentalphysik, Luruper Chaussee 149, D 22761 Hamburg, Germany \and
Laboratoire Univers et Particules de Montpellier, Universit\'e Montpellier 2, CNRS/IN2P3,  CC 72, Place Eug\`ene Bataillon, F-34095 Montpellier Cedex 5, France \and
Max-Planck-Institut f\"ur Kernphysik, P.O. Box 103980, D 69029 Heidelberg, Germany \and
Dublin Institute for Advanced Studies, 31 Fitzwilliam Place, Dublin 2, Ireland \and
National Academy of Sciences of the Republic of Armenia, Yerevan  \and
Yerevan Physics Institute, 2 Alikhanian Brothers St., 375036 Yerevan, Armenia \and
Universit\"at Erlangen-N\"urnberg, Physikalisches Institut, Erwin-Rommel-Str. 1, D 91058 Erlangen, Germany \and
Nicolaus Copernicus Astronomical Center, ul. Bartycka 18, 00-716 Warsaw, Poland \and
CEA Saclay, DSM/IRFU, F-91191 Gif-Sur-Yvette Cedex, France \and
University of Durham, Department of Physics, South Road, Durham DH1 3LE, U.K. \and
Astroparticule et Cosmologie (APC), CNRS, Universit\'{e} Paris 7 Denis Diderot, 10, rue Alice Domon et L\'{e}onie Duquet, F-75205 Paris Cedex 13, France \thanks{(UMR 7164: CNRS, Universit\'e Paris VII, CEA, Observatoire de Paris)} \and
Laboratoire Leprince-Ringuet, Ecole Polytechnique, CNRS/IN2P3, F-91128 Palaiseau, France \and
Institut f\"ur Theoretische Physik, Lehrstuhl IV: Weltraum und Astrophysik, Ruhr-Universit\"at Bochum, D 44780 Bochum, Germany
% ...
}

\date{Received 17 June 2011; accepted 16 January 2012}
 
 \abstract
   {Multiwavelength (MWL) observations of the blazar PKS~2155-304 during two weeks in July and August 2006, the period when two exceptional flares at very high energies (VHE, E$\gtrsim$ 100 GeV) occurred,  provide a detailed picture of the evolution of its
   emission. The complete data set from this campaign is presented, including observations in VHE $\gamma$-rays (H.E.S.S.), X-rays ({\it RXTE}, {\it CHANDRA}, \emph{SWIFT} XRT), optical ({\it SWIFT} UVOT, Bronberg, Watcher, ROTSE), and in the radio band (NRT, HartRAO, ATCA). 
   Optical and radio light curves from 2004 to 2008 are compared to the available VHE data from this period, to put the 2006 campaign into the context of the long-term evolution of the source. \thanks{Tables corresponding to Figures~\ref{fig:multilambda} and~\ref{fig:lc_radio_rotse_hess} are only available in electronic form at the CDS via anonymous ftp to cdsarc.u-strasbg.fr (130.79.128.5) or via http://cdsweb.u-strasbg.fr/cgi-bin/qcat?J/A+A } }
   {The data set offers a close view of the evolution of the source on different time scales and yields new insights into the properties of the emission process. The predictions of synchrotron self-Compton (SSC) scenarios are compared to the MWL data, with the aim of describing the dominant features in the data down to the hour time scale.}
   { The spectral variability in the X-ray and VHE bands is explored and correlations between the integral fluxes at different wavelengths are evaluated. SSC modelling is used to interpret the  general trends of the varying spectral energy distribution.}
   {The X-ray and VHE $\gamma$-ray emission are correlated during the observed high state of the source, but show no direct connection with longer wavelengths. The long-term flux evolution in the optical and radio bands is found to be correlated and shows that the source reaches a high state at long wavelengths after the occurrence of the VHE flares. Spectral hardening is seen in the \emph{SWIFT} XRT data.} 
 {The nightly averaged high-energy spectra of the non-flaring nights can be reproduced by a stationary one-zone SSC model, with only small variations in the parameters.
 The spectral and flux evolution in the high-energy band during the night of the second VHE flare is modelled with multi-zone SSC models, which can provide relatively simple interpretations for the hour time-scale evolution of the high-energy emission, even for such a complex data set. For the first time in this type of source, a clear indication is found for a relation between high activity at high energies and a long-term increase in the low frequency fluxes.}


 \maketitle


\section{Introduction}

Several aspects of the exceptionally high state of the blazar PKS~2155-304 that was observed by the Imaging Air Cherenkov Telescopes (IACT) of the High Energy Stereoscopic System (H.E.S.S.) during the summer of 2006 have already been presented in \cite{vhe_paper} and \cite{Aha2009b}. These publications focused on the spectral and temporal 
variability in the very high energy (VHE, E$\gtrsim$100 GeV) band and on the short-term multiwavelength (MWL) behaviour during the night of the second VHE flare (hereafter ``Flare 2'';  MJD 53946). The present paper completes this work by providing a long-term MWL view from radio to VHE during the whole of the 2006 campaign and beyond. The MWL behaviour of the source during the 2006 high state is analysed with stationary and time-dependent emission models and put in the context of the long-term data set. 

Obtaining a multispectral view of the temporal evolution of blazar fluxes is one of the main keys towards a global understanding of AGN physics.
MWL observations help to gain detailed insight into the acceleration and emission processes of relativistic particles, which are thought to occur close to the central
black hole \citep[e.g.][]{Ner2007, Rie2008, Sol2009} or in the jet \citep[e.g.][]{Sik1994, Ino1996,Kat2001, Kat2003, Tav2008}. The double-bump structure of the spectral energy distribution observed in blazars at high energies is interpreted differently in leptonic and hadronic models, which suppose that the particle population in the 
source is dominated by either electrons and positrons or by hadrons. 

The first bump, which is found at X-ray/UV energies in high-frequency peaked BL Lacs (HBLs) and at optical energies in 
low-frequency peaked BL Lacs (LBLs) and flat-spectrum radio quasars (FSRQs), is usually ascribed to synchrotron emission from a relativistic population of electrons. The second bump, at GeV/TeV energies in HBLs and usually at hard X-rays in LBLs and FSRQs, is interpreted as inverse Compton (IC) emission from electrons up-scattering synchrotron or external photons in leptonic models. In the hadronic scenarios, it arises instead from proton synchrotron emission or from hadronic interactions inside the jet or between the jet outflow and the ambient medium.  

For HBLs, external photon fields are generally assumed to play a minor role in the emission, since no strong emission lines are observed. Thus the synchrotron self-Compton (SSC) scenario, 
where the bump at highest energies is attributed to synchrotron photons that have been up-scattered by their parent electron population, is often applied to interpret spectral energy distributions 
from these objects. SSC models require only a very restrained set of free parameters, compared to those leptonic scenarios that add an external component or compared to the more complex hadronic models. Throughout the modelling presented in this paper, the SSC approach will be adopted and the term ``one-zone model'' will be applied to models where both the X-ray and $\gamma$-ray components are largely dominated by emission from the same zone. 

If one wants to put constraints on the different existing scenarios, the active states of blazars provide crucial information on the characteristics of the emission region and the emission process, since they can exhibit high variability on short time scales and over a wide wavelength range.

PKS~2155-304 is one of the brightest BL Lacs in the X-ray \citep[e.g.][]{Bri1994, Kub1998, Gio1998, Ves1999} and EUV \citep{Mar1993} bands. 
The source is classified as an HBL and has a redshift of $z=0.116$.  
Since the discovery of X-ray emission from this object \citep{Gri1979, Sch1979}, it has been repeatedly observed over a wide range of frequencies from radio to 
VHE $\gamma$-rays \citep[e.g.][]{Tre1989, Ede1995, Ves1995, Zha1996, Urr1997, Pia1997, Pin2004, Dom2004, Aha2005a, Zha2006, Ost2007}. BeppoSAX
observations from 1996 to 1999 show X-ray variability on time scales of $\approx$ 1 hour \citep{Zha2002}. 
Several physical implications for the emission mechanisms have been reported \citep[e.g.][]{Chiap1999, Katao2000, Ede2001, Tan2001, Zha2002, Zha2006}.

In November 1997, an outburst from PKS~2155-304 was detected in X-rays and $\gamma$-rays with {\it EGRET} \citep{Sre1997}, {\it BeppoSAX} \citep{Chiap1999}, and {\it RXTE} \citep{Ves1999}. During this active
phase, the Durham group reported the first detection of VHE $\gamma$-rays at a level of 6.8 standard deviations ($\sigma$) above 300 GeV \citep{Cha1999a, Cha1999b}. In 1998, when the X-ray
flux level was low, they
had not found any evidence of TeV $\gamma$-ray emission \citep{Cha1999}.  PKS~2155-304 was also observed with the CANGAROO-I 3.8m telescope in 1997. No $\gamma$-ray signal above 1.5 TeV was detected \citep{Rob1999}. The source was further observed in 1999, 2000, and 2001 with the CANGAROO-II telescope. It remained in a low state of
X-ray activity in those periods, and was not detected above the energy threshold of 420 GeV \citep{Nis2001, Nis2002, Nak2003}. 

PKS~2155-304 was confirmed as a TeV $\gamma$-ray source by the H.E.S.S. collaboration with observations in 2002 and 2003. A detection with a significance of 45 $\sigma$ at energies greater than 160 GeV was reported \citep{Aha2005a}. 
A first MWL campaign of PKS~2155-304 including H.E.S.S. data was conducted in 2003 over several weeks, during the construction phase of the array \citep{Aha2005b}. The source was observed 
simultaneously in the X-ray range by {\it RXTE}/PCA, in the optical by ROTSE and in the radio band by the Nan\c{c}ay Radio Telescope (NRT) and was found to be in a low state. Intra-night variability was seen in the 
VHE and X-ray band, with the shortest time scales detected with {\it RXTE} $\approx$25 minutes. The optical flux showed only moderate variation. No correlation was observed between the different bands. 
The hardness ratio in X-rays showed an increase in the spectral hardness with higher flux levels. Data from another MWL campaign conducted by the H.E.S.S. collaboration in 2004 were affected by the poor quality of the atmosphere at the time of data acquisition.

The most recent MWL observations of the source in 2008, including data from H.E.S.S., ATOM, {\it RXTE} and for the first time data in the high-energy $\gamma$-ray range from {\it Fermi}, have been reported in \cite{Aha2009a}. The source was found with a relatively low VHE flux level and no correlation between VHE $\gamma$-rays and X-rays was observed. There was, however, some evidence of correlated behaviour between the VHE and optical bands.

The observation of a spectacular flare  (hereafter "Flare 1") on July 28, 2006 (MJD 53944), with a more than 20-fold flux increase in one night and variability time scales down to 200 seconds was reported in \cite{Aha2007}. Observations with Cangaroo-III confirmed the detection of a very active state with variability on sub-hour time scales \citep{Sak2008}. The very fast variability seen by H.E.S.S.
implies in the one-zone SSC framework that the Doppler factor\footnote{$ \delta_\mathrm{b} = \left[ \Gamma_\mathrm{b} \left( 1 - \beta_\mathrm{b} \cos{ \theta} \right)  \right]^{-1}$, where $\beta_\mathrm{b}$ is the velocity of the emission region in c units, $\Gamma_\mathrm{b}$ the bulk Lorentz factor, and $\theta$ the angle between the jet axis and the line of sight.} 
$\delta_\mathrm{b}$ had a value of about 60 to 120 times the size of the emission region in units of the Schwarzschild radius of the central black hole (see also \cite{Beg2008}). This means that either the Doppler factor was unusually large compared to previous blazar observations or that the emission region was much smaller than the size of the central black hole. 

A study of simultaneous observations in the X-ray and VHE bands during Flare 2 \citep{Aha2009b} found a very steep correlation between the X-ray flux detected by {\it CHANDRA} and the
VHE $\gamma$-ray flux, with the flux in the VHE band decreasing approximately as the third power of the X-ray flux.  No indication for a time-lag between those bands was detected. As a consequence of the steep correlation, one-zone SSC models are strongly disfavoured to describe the emission in X-rays and VHE  $\gamma$-rays during this flare \citep[for a general discussion see][]{Kat2005}.  

A further investigation of the emission mechanism in the VHE band during the four most active nights, MJD 53944 to 53947, has led to the conclusion that the observed flux variability stems from a lognormal stochastic process \citep{vhe_paper}. The existence of a quiescent state of the source and of complex spectral variability in the VHE band was equally demonstrated by 
combining the data from 2006 with older and newer data sets.

In the following, the complete two-week MWL campaign of 2006 is described and an interpretation of the observed flux evolution in the different energy bands is given. The evolution of the VHE flux and spectrum of PKS~2155-304 on different time scales, covered by \cite{vhe_paper}, is here compared to the MWL emission from the radio to the X-ray band. The physical processes behind Flare 2 were discussed in general terms by \cite{Aha2009b}. Here, the data from this event are put into the context of the MWL observations before and after the flare and are confronted with actual emission models. A comprehensive summary of all the available MWL data that correspond to the period of the H.E.S.S. campaign 
is provided in Section~\ref{section:observations}. In Section~\ref{section:results}, a study of correlations between light curves at different wavelengths is presented and
the spectral variability in the X-ray and VHE band is described. Modelling of the nightly averaged spectral energy distributions with a stationary one-zone SSC model is presented in Section~\ref{section:modelling}. For the night of Flare 2, the only night with simultaneous H.E.S.S. and {\it CHANDRA} coverage, three different time-dependent SSC models are discussed.  The insight gained from these models and from the MWL analysis is discussed in Section~\ref{section:discussion}. No attempt is made 
at describing the very rapid variability observed in the H.E.S.S. data on the time scale of only a few minutes, which in itself poses a challenging problem. This issue is discussed briefly in
Section~\ref{section:discussion}.

Throughout the paper, a  flat cosmology with $H_0 =$ 70 km s$^{-1}$  Mpc$^{-1}$, $\Omega_\mathrm{M} =$ 0.3, and $\Omega_\mathrm{{\lambda} }=$ 0.7 is assumed. Unless otherwise 
indicated, all errors are statistical and are given at the 1$\sigma$ confidence level for one parameter of interest. Dates of observations are in general given in the Modified Julian Day
(MJD) format.

\section{Observations and data analysis}
\label{section:observations}

 \begin{figure*}
     \centering
     \includegraphics[width=24 cm, angle=90]{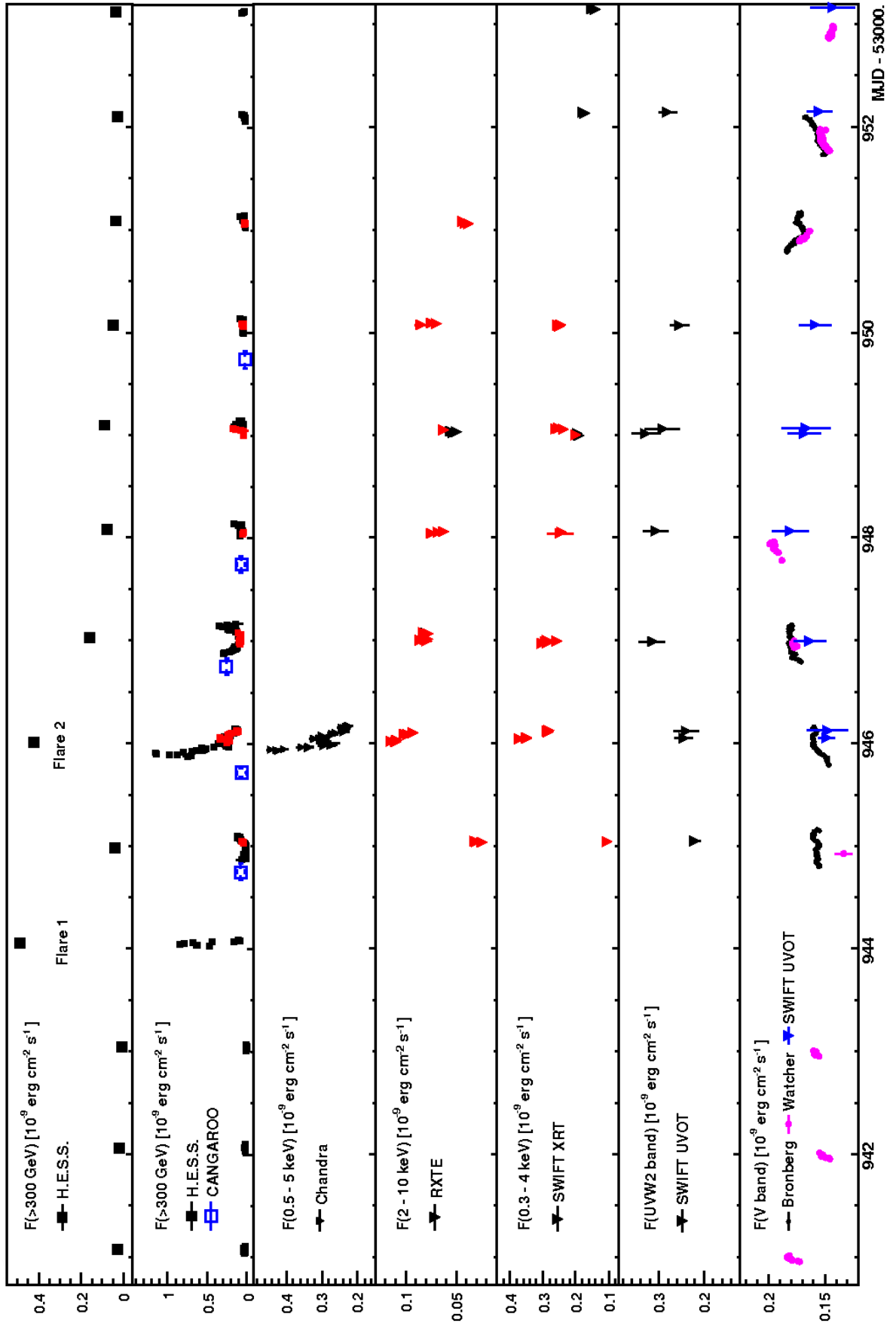}
  \caption{Light curves in the VHE, X-ray, UV and optical bands during the MWL campaign. Details on the different data sets are given in the text. "Flare 1" and "Flare 2" refer to 
  the two exceptional VHE flares seen before and during the MWL campaign. All light curves are binned in intervals of 10 minutes, except for the nightly averaged H.E.S.S. fluxes (first panel) and the averaged CANGAROO fluxes (second panel). The published flux from CANGAROO has a threshold of 660 GeV. It has been extrapolated down to 300 GeV for comparison with H.E.S.S.  The H.E.S.S. points marked in red correspond to nights where simultaneous data are available either from {\it RXTE} or from {\it SWIFT} XRT. The corresponding simultaneous points are also marked in red in the light curves from the latter. The {\it SWIFT} UVOT flux densities \citep{Fos2008} in the UVW2 and v bands have been multiplied by 1928 $\AA$ and 5500 $\AA$, respectively. The UVOT error bars include systematic errors.} 
   \label{fig:multilambda}
    \end{figure*}

\subsection{VHE $\gamma$-ray data}
\label{subsec:vhedata}

Due to its high brightness in the VHE range, PKS~2155-304 is a regular target for observations with H.E.S.S. and has been followed every year since the start of operations. Scheduled for observations
towards the end of July 2006, the blazar was found in a state of high activity on the first night of the observational period, the night of July 25/26 (MJD 53941), with a significance of 35$\sigma$ (standard deviations) for 1.31 hours of live time, which is on average 89\% of the flux from the Crab nebula as observed by H.E.S.S. above 200 GeV. On the following night, the source was seen with 29.5$\sigma$ for 1.76 hours of live time. On the night of July 27/28 (MJD 53943), the night before the first flare, the source was at a lower, but still relatively high flux level, at 12.8$\sigma$ for 1.33 hours of live time. This exceptional activity triggered Target of Opportunity observations on the source with SWIFT and RXTE; the monitoring however started only one day after the major flare of July 28/29 (MJD 53944), where PKS~2155-304 was found at 172.9$\sigma$ for 1.33 hours of live time.

The source remained in a relatively high state during the whole campaign, with fluxes on most nights exceeding the quiescent state, derived from data taken between 2005 and 2007 \citep{vhe_paper}, by an order of magnitude. The two flares detected on the nights of the 28th and 
30th of July 2006 reached peak fluxes close to two orders of magnitude above the quiescent state. The H.E.S.S. light curves for all the nights of the 2006 observations are included in Figure~\ref{fig:multilambda} together with light curves from the other observed wavelength bands, excluding the long-term optical and radio data described in Sections~\ref{subsec:optical} and~\ref{subsec:radio}.

Details of the observations and data analysis of the 2006 H.E.S.S. data have been given by~\cite{vhe_paper}. To extract light curves from the H.E.S.S. data, the VHE $\gamma$-ray flux was integrated above 200 GeV or above the (zenith-angle dependent) analysis threshold, if the latter was higher. After extraction of the spectral shape for each night, integrated fluxes were then determined above 300 GeV. 

  \begin{figure*}
     \centering
      \includegraphics[width=18 cm]{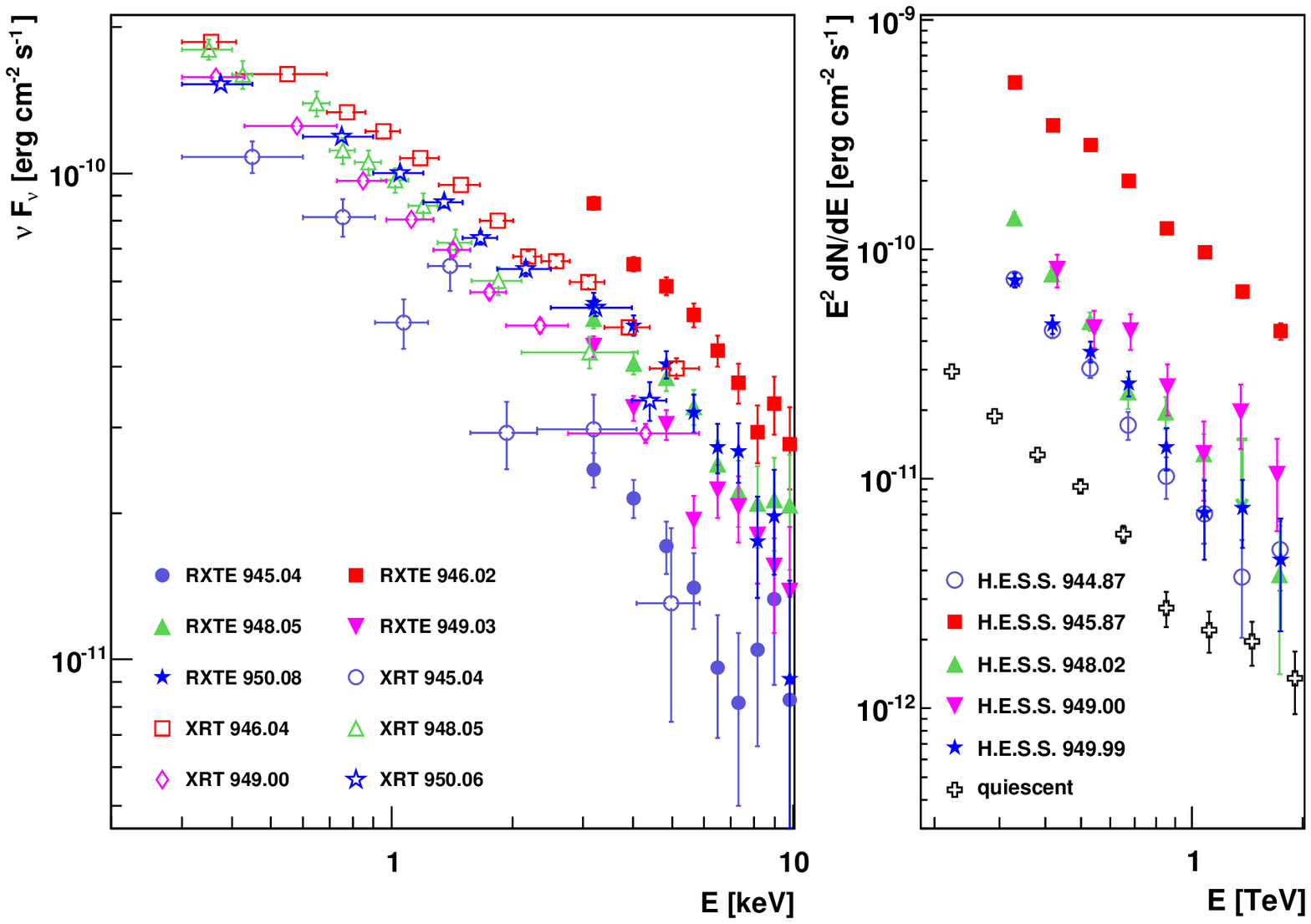}
  \caption{ Left panel: spectra obtained with {\it SWIFT} XRT and  {\it RXTE} during nights of the 2006 MWL campaign. Only nights where spectra for both instruments could be extracted are included. For MJD 53946 only the {\it RXTE} spectrum from the first pointing and for MJD 53947 only the spectrum
  from the second {\it RXTE} pointing have been included to avoid cluttering the plot. The spectra have been corrected for Galactic absorption.
  Right panel: nightly averaged H.E.S.S. spectra for the same selection of nights. Upper limits at 99\% confidence level are indicated as arrows. No correction for absorption on the extra-galactic background has been applied. The quiescent flux is taken from~\cite{vhe_paper}.  The legends in both panels provide the start times of the observations in MJD - 53000.}
     \label{fig:spec_rxtenights}
    \end{figure*}

The 2006 MWL data set includes seven nights with {\it RXTE} X-ray pointings and eight nights with {\it SWIFT}-XRT X-ray pointings, which were performed during H.E.S.S. observations. Unfortunately, no MWL coverage exists 
for Flare 1. The H.E.S.S. spectra for the nights with X-ray coverage, extracted between 0.3 and 2 TeV based on the hypothesis of a power law, are shown in Figure~\ref{fig:spec_rxtenights}, with the corresponding photon indices listed in Table~\ref{tab:vhe} in Appendix~\ref{sec:app0}. The spectrum of the quiescent state is included for comparison. The spectra have been determined from the measured distributions of
signals from the source and from an off-source region using a forward folding maximum-likelihood method (for more details cf. \cite{vhe_paper}). The forward folding method yields the best set of parameters
for the power law hypothesis with their associated errors and the residuals between the measured and expected excess in each energy bin. To allow an easier comparison with the other wavelength bands, spectral points have been determined from the residuals. The $\chi^2$ values given in Table~\ref{tab:vhe}, not provided directly by the maximum-likelihood method, have been determined from power law fits to the spectral points.
The power law yields a good description for all nights, except for the data set from Flare 2 (starting at MJD 53945.87), as discussed in depth by \citet{Aha2009b}. For the night of Flare 2, the power law merely provides a rough approximation of the average spectrum, which is still useful for a comparison with the other nights.

It should be noted that the H.E.S.S. data during MJD 53949 were taken with only two telescopes. Nevertheless, the observed events were of sufficient quality to be used for the extraction of a spectrum for this night, although systematic uncertainties are larger than for the other nights. Given the higher threshold for observations with only two telescopes ($\sim$ 420 GeV in this case), the integral flux above 300 GeV for this night had to be determined using an extrapolation of the spectrum. Data from the three nights MJD 53951 to MJD 53953, where the source was in a relatively low state and exposure times were short, have been averaged to derive a spectrum with good statistics.

The nightly averaged VHE flux varies by about an order of magnitude for the nights with simultaneous X-ray data. The comparison with the spectrum of the quiescent state shows that all nights of the 2006 campaign have a significantly elevated flux level.

\subsection{X-ray data}
\label{subsec:xraydata}

The exceptional activity detected by H.E.S.S. triggered an {\it RXTE} observation on the source; the monitoring started only one day after Flare 1, on MJD 53945.
The PCA \citep{Jah1996} units of the {\it RXTE} telescope observed the source with exposures of typically $\approx700\,\rm s$ per pointing. One pointing was performed per night, except for two nights (MJD 53946 and 53947) with two pointings each.The STANDARD2 data were extracted using the {\tt HEASOFT 6.5.1} analysis software package provided by NASA/GSFC, and filtered using the {\it RXTE} Guest Observer Facility (GOF) recommended criteria. The spectra were extracted using {\tt XSPEC v.12.4.0}, with a fixed column density of $N_{\rm H} = 1.7\times 10^{20}\,\rm cm^{-2}$ \citep{Dic1990}. A power law model was used for spectral fitting. No significant improvement was noted when using a broken power law.

A comparison of the  {\it RXTE}  energy spectra is shown in Figure~\ref{fig:spec_rxtenights} with the results of a power law fit given in Table~\ref{tab:rxte} in Appendix~\ref{sec:app0}. 
The corresponding light curves, binned in time intervals of 10 minutes, are included in Figure~\ref{fig:multilambda}. 
The lowest X-ray fluxes are close to the level detected in the 2003 MWL campaign, when the source was in a rather low state and the integral flux between 2 and 10 keV was measured at 2.66 $\times$ 10$^{-11}$ erg cm$^{-2}$ s$^{-1}$ \citep{Aha2005b}.

Simultaneous data from the {\it CHANDRA} telescope were only available for the night of Flare 2. The decreasing part of an X-ray flare is visible during
that night, occurring simultaneously with the VHE flare. The light curve (integrated flux between 0.5 and 5 keV) for this night, rebinned in 10 min intervals, is included in 
Figure~\ref{fig:multilambda}. A detailed description of those data and the analysis is published by~\cite{Aha2009b}.  

Data taken with XRT onboard {\it SWIFT} during the H.E.S.S. 2006 campaign show an X-ray flare during the night
of Flare 2, which then decreased by a factor of $\approx 5$ in a month \citep{Fos2007}.
The light  curves of {\it SWIFT} XRT and {\it SWIFT} UVOT are included in Figure~\ref{fig:multilambda}, rebinned in 10 minute bins.  
{\it SWIFT} XRT data are available for seven nights, six out of which coincide with nights where {\it RXTE} pointings were performed as well.

The {\it SWIFT} XRT data for the nights during the H.E.S.S. campaign have been analysed in the present work assuming a column density $N_{\rm H} = 1.7 \times 10^{20} \mathrm{cm}^{-2}$, to be directly comparable with the {\it CHANDRA} 
and {\it RXTE} data. This analysis has been carried out with the {\tt HEASOFT 6.5} standard tools. The data were binned requiring a minimum of 20 counts per bin and fitted to a single or broken power law. 
Only points in the intervals from 0.3 to 0.45 keV and from 0.6 to an upper limit of between 4 and 7 keV (depending on the data set) were included in the fit to suppress known systematic effects at intermediate energies. 
As \cite{Fos2007} have already pointed out, the spectra of several nights are best fit by a broken power law. In the new analysis, this was the case for all nights except for MJD 53945, 53952 and 53953, where a broken power law did not yield a better result than a single power law. During those three nights, the duration of the pointings was significantly shorter than for the other nights and sufficient photon statistics are only available up to about 4 keV. The data recorded during the night of MJD 53947 do not provide sufficient statistics to extract a spectrum. The resulting photon indices and de-absorbed fluxes are listed in Table~\ref{tab:xrt} in Appendix~\ref{sec:app0}, together with the start times and the durations of the pointings.  

\subsection{Optical and UV data}
\label{subsec:optical}

PKS~2155-304 was observed with several telescopes in the optical and UV range during the H.E.S.S. campaign. Even though the coverage of the nights of the campaign is not complete,
a compilation of the different available data sets provides a good picture of the behaviour of the source in these bands.
The optical flux from the host galaxy is estimated to have an apparent R magnitude of 15.1 \citep{Aha2005b}, based on measurements in the optical and near infrared \citep{Fal1996, Kot1998}. This contribution is not significant for a discussion of the flux variation and is therefore neglected in the following.

\begin{figure*}
     \centering
      \includegraphics[width=18cm]{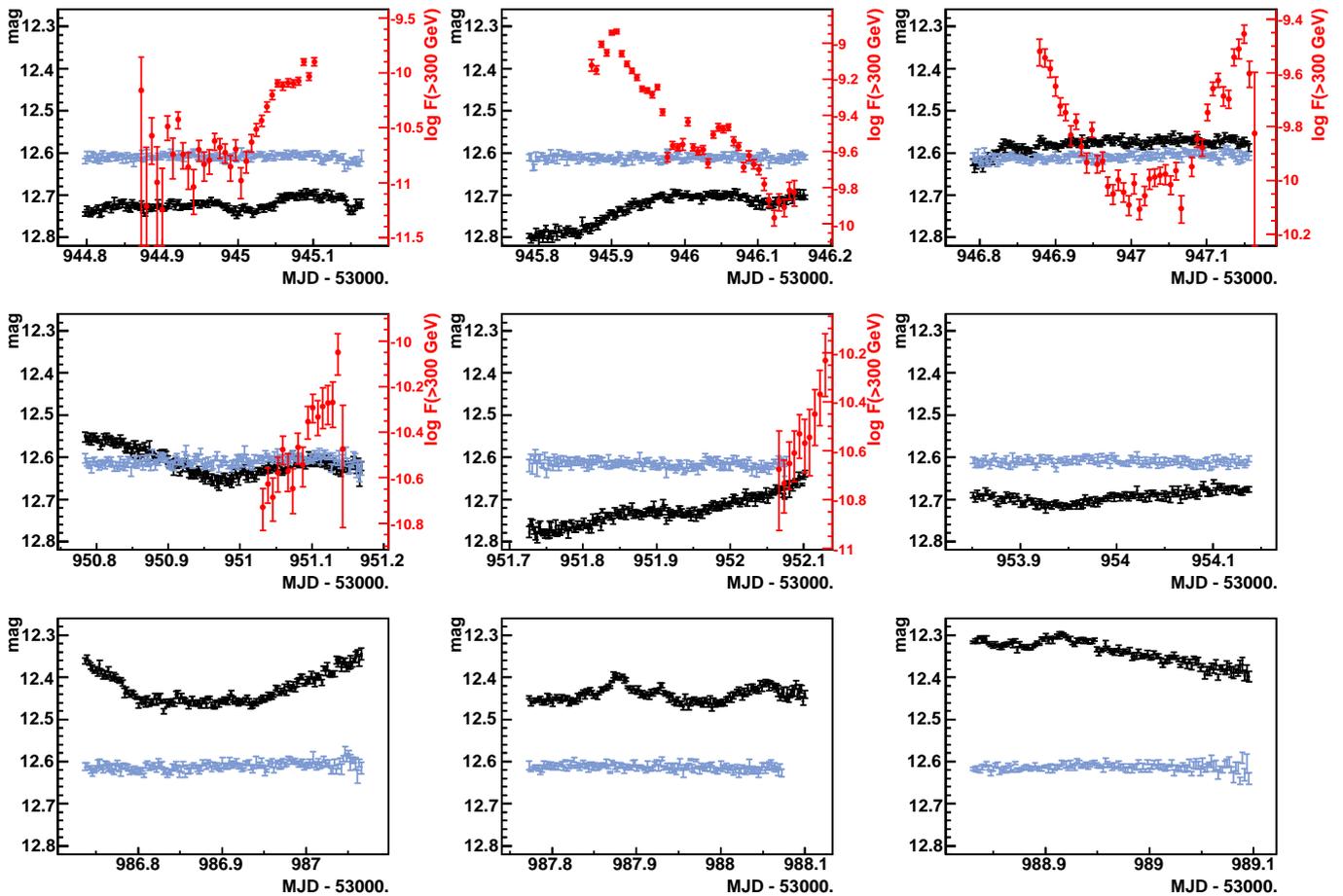}
    \caption{From the upper left to the lower right: PKS~2155-304 light curves from Bronberg optical data (black points) for the five nights of observation that coincided with the H.E.S.S. campaign in 2006 (MJD 53945, 53946, 53947, 53951, 53952),
    for the night MJD 53954 and for three nights in September (MJD 53987 to 53989). The flat light curve of a reference star is shown for comparison (light blue points). For the nights with simultaneous H.E.S.S. coverage, the VHE light curve is shown for comparison (red points, y-axis on the right). The data have not been corrected for Galactic absorption.}
    \label{fig:lc_bronberg}
 \end{figure*}

For five of the nights with H.E.S.S. observations, data in the optical range (V band) are available from the South African Bronberg observatory~\citep{Mon2007}. Observations were made with a 35 cm f/8 ``Meade'' telescope.
Calibration of the frames was obtained by calculating magnitude shifts relative to two bright isolated stars in the field.
The error on the average is less than 0.015 mag. The first star was used as a reference and its very stable light curve is shown together with the light curve obtained from PKS~2155-304 in Figure~\ref{fig:lc_bronberg}. 

In addition to the observations that were simultaneous with the H.E.S.S. campaign, light curves from one night in August 2006 and from three nights in September 2006 were available and have been included in this figure. For all light curves, series of six successive data points were averaged to yield a mean value and error of the mean over a time period of approximately 151 seconds. This bin size was chosen, after comparison with the signal from the reference star, to average out fluctuations in the data induced mostly by variability in the atmosphere. 
 
Optical and UV data from {\it SWIFT} UVOT are also available for the H.E.S.S. observational period \citep{Fos2007}. The UV flux during the H.E.S.S. campaign is larger by a factor of about 1.5 compared to data from April of 2006.

\begin{figure}
     \centering
        \resizebox{\hsize}{!}{\includegraphics[width=\columnwidth]{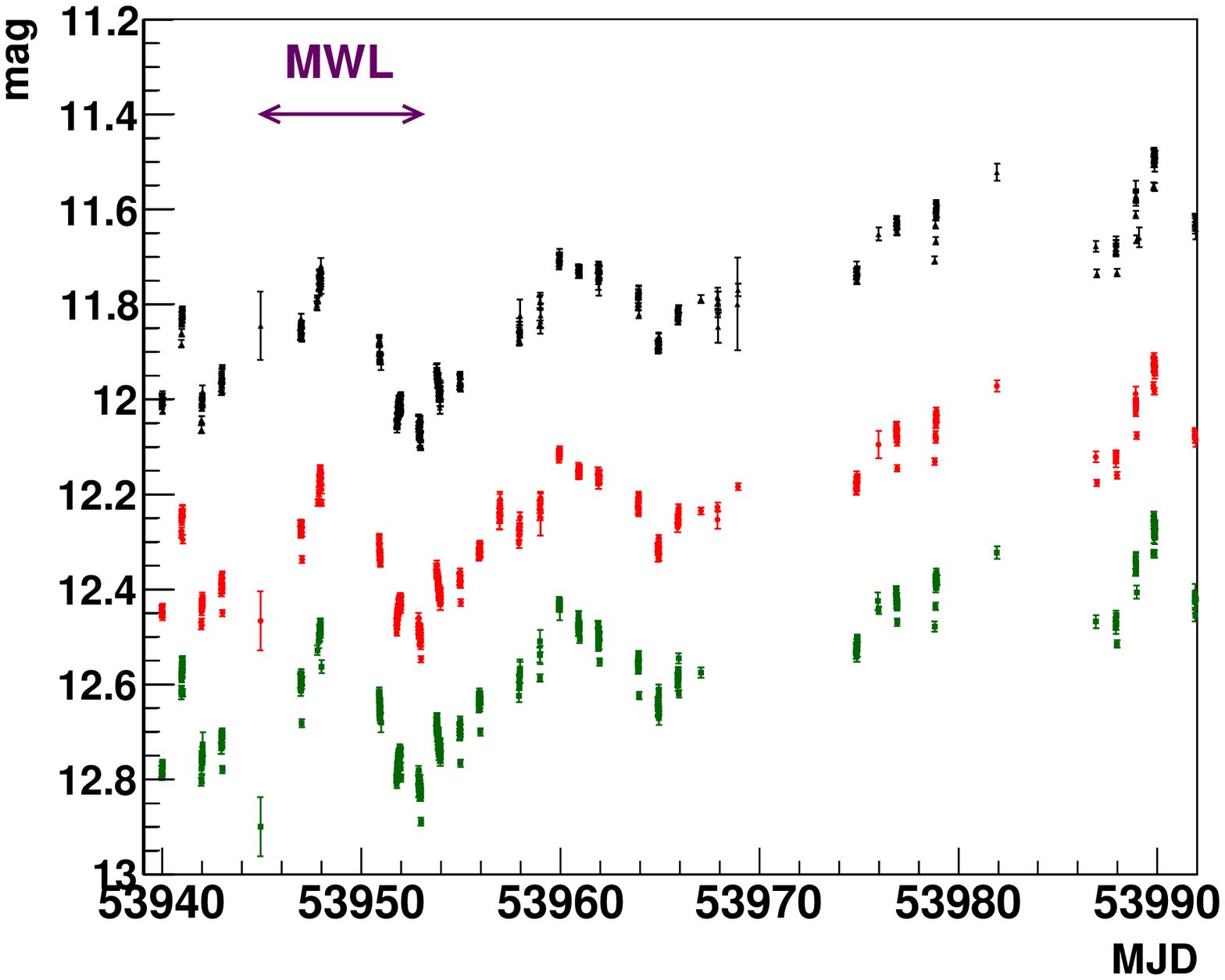}}
    \caption{Light curves of Watcher optical data in the I band (upper points; in black), R band (points in the middle; in red)
     and V band (lower points; in green). The arrow indicates the time interval of the MWL campaign. The data have not been corrected for Galactic absorption.}
    \label{fig:lc_watcher}
 \end{figure}

Light curves from the Watcher telescope, situated in the South African Boyden Observatory\footnote{website: http://www.assabfn.co.za/friendsofboyden/boyden.htm }, are shown in Figure~\ref{fig:lc_watcher} for the R, V and I optical band. Watcher is a robotically
controlled 40 cm f/14.25 classical Cassegrain telescope.
The flux from PKS~2155-304 has been calibrated against at least two and up to five reference stars in the field of view. The uncertainty in each point of the light curve is estimated as the RMS of the fluctuations around the average fluxes from the reference stars. A rotation of the CCD camera by 180$^{\circ}$ close to midnight during each night led to a systematic variation in the brightness of the reference stars by up to 6\%. A residual effect on the
light curves of PKS~2155-304 could not be completely corrected in the analysis, which led to discrepancies with the Bronberg light curves for Watcher data taken after the rotation. It was decided to exclude all data points for each night after the rotation occurred. After this correction, the Watcher and Bronberg light curves are found to be in very good agreement. 

 \begin{figure}
     \centering
     \vspace{0.1 cm}
      \resizebox{\hsize}{!}{\includegraphics[width=\columnwidth]{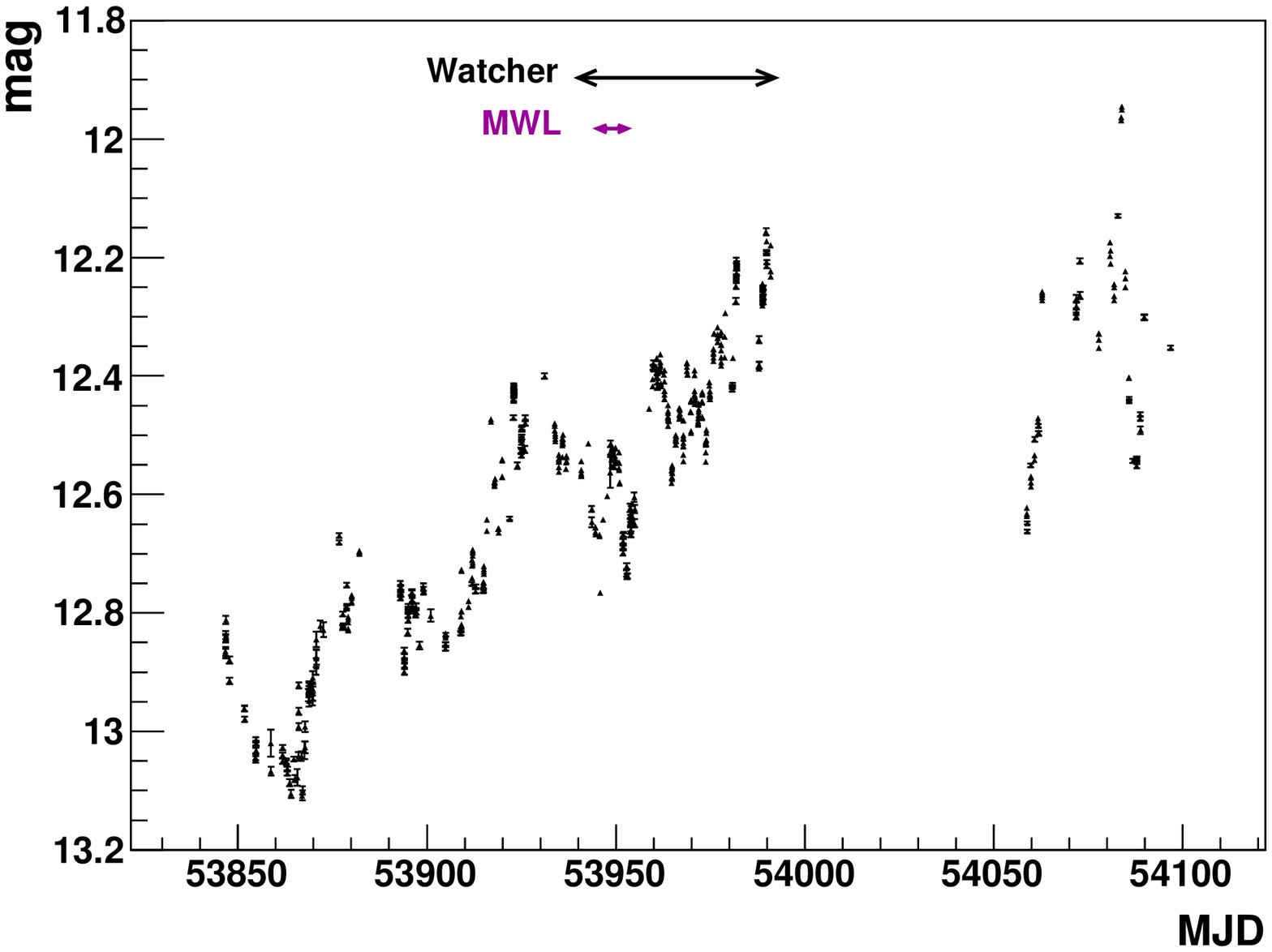}}
    \caption{PKS~2155-304 light curve observed with the ROTSE telescope in 2006 (R band magnitude). The arrows indicate the time intervals of the MWL campaign and of coverage with the
    Watcher telescope. The data have not been corrected for Galactic absorption.}
    \label{fig:lc_rotse}
 \end{figure}

Coverage of the source over a longer time period was carried out with the robotically controlled ROTSE-IIIc telescope, located on the H.E.S.S. site. The resulting light curve for all data taken in 2006 is shown in
Figure~\ref{fig:lc_rotse}. The complete data set of available ROTSE data from April 2004 to November 2008 is included in 
Figure~\ref{fig:lc_radio_rotse_hess} and will be discussed in Section~\ref{section:results}. ROTSE has a wide field of view (1.85$^{\circ}~ \times$ 1.85$^{\circ}$) and is operated without filters.
A relative R magnitude is derived by comparison of the instrumental 
magnitude with the USNO catalogue as described by~\cite{Ake2000}. A detailed description of the system is given by~\cite{Ake2003} and more
information on the data analysis can be found in~\cite{Aha2005b}. During the period where simultaneous Watcher data are available, the form of the ROTSE and Watcher light curves are in good 
agreement, although the normalization of the ROTSE data is lower by about 0.2 mag due to known systematic biases in the ROTSE data.
In 2003, during a low state of the source, an optical flux between 13.3 and 13.7 in relative R magnitude was measured by ROTSE. Figure~\ref{fig:lc_rotse} shows a higher
optical activity and a clear increase of the optical flux in 2006.

\subsection{Radio data}
\label{subsec:radio}

\begin{figure*}
     \centering
      \includegraphics[width=18 cm]{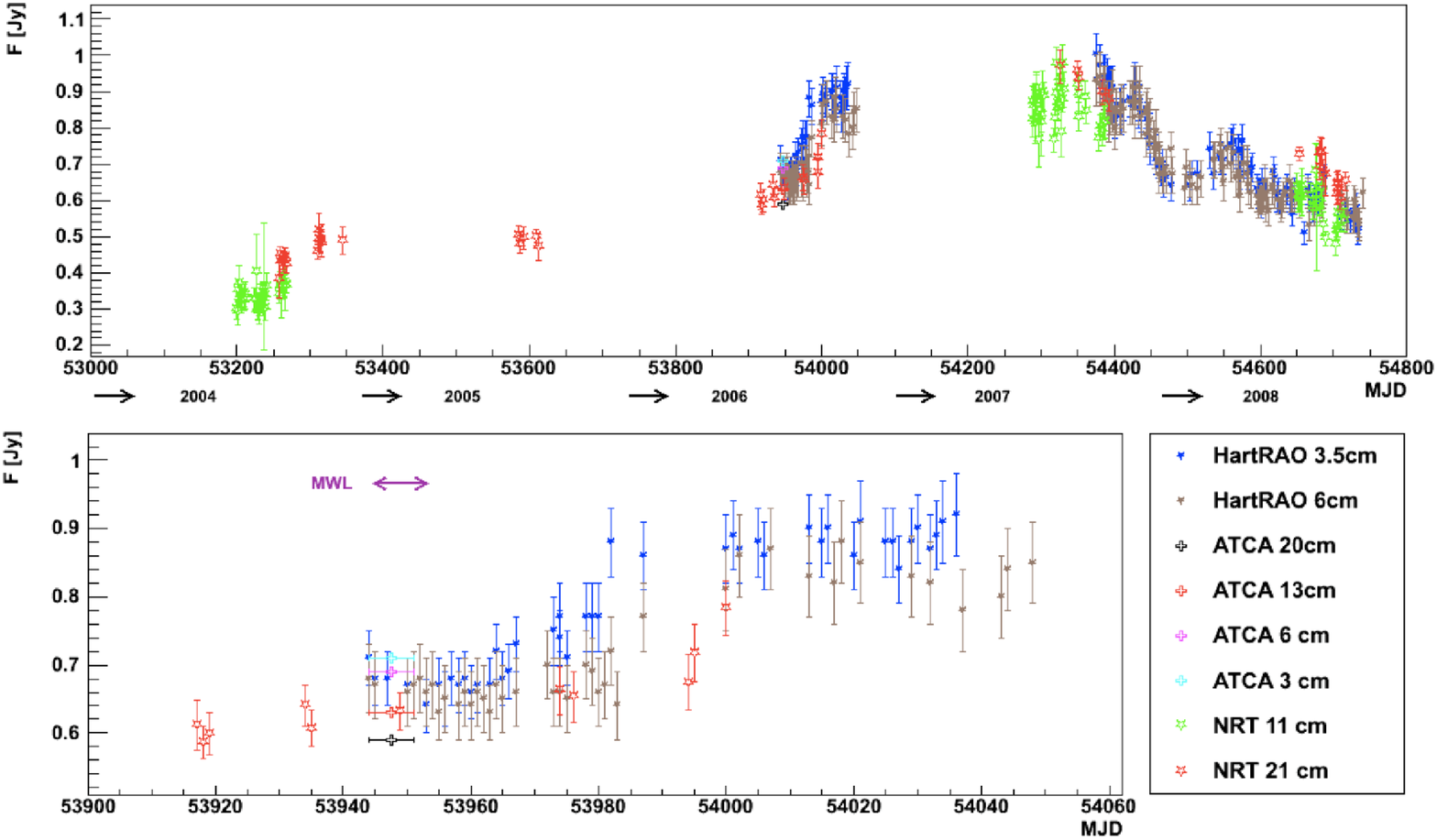}
  \caption{Long term radio light curve with data from the NRT, ATCA and HartRAO, from July 2004 to November 2008. The lower panel is a zoom on the period around the H.E.S.S. campaign. HartRAO error bars include an estimate of the systematic errors. The double arrow marked ``MWL'' in the lower panel indicates the time period of the MWL campaign.}
     \label{fig:lc_radio}
    \end{figure*}

Radio data from three observatories are available for the period of the 2006 H.E.S.S. campaign, as well as for a longer interval of several months and years before and after the campaign. 
A long-term light curve including measurements at different frequencies is shown in Figure~\ref{fig:lc_radio}. 

A monitoring programme with the Nan\c{c}ay Radio Telescope (NRT) of extragalactic sources observed with H.E.S.S. is in place since 2001. 
The NRT consists of a single-dish antenna with a collecting area of 200 $\times$ 34.56 m$^2$.
Observations of PKS~2155-304 were made at a wavelength of 11 cm, with a half-power beam-width (HPBW) of $\sim$ 2' (RA)$ \times$ 10' (Dec.) \citep{The2007}, and at 21 cm, with HPBW $\sim$ 4' $\times$ 22'. 
Each observation was based on between 2 and 36 one-minute drift scans at frequencies of 2679 and 2691 MHz, as well as 1407 and 1420 MHz. The flux was calibrated against the stable radio source 4C~55.16 for the period from MJD 53558 to MJD  54100, covering the H.E.S.S. campaign. The calibrators 3C~123, 3C~161, 3C~286 and 3C~295 were used for the data analysis before and after this period. 

The Hartebeesthoek Radio Astronomy Observatory (HartRAO), located in South Africa, is equipped with a 26 m diameter single dish\footnote{website: http://www.hartrao.ac.za }.
Observations with HartRAO started on July 28, after an alert from the H.E.S.S. collaboration. Between 3 and 5 scans were performed each day for the following two weeks.
Flux measurements at 6 cm (HPBW $\sim$ 10') and 3.5 cm (HPBW $\sim$ 6') were carried out with dual-beam receivers. Continuous flux measurements of PKS~2155-304 used the drift scan method, with scans offset at the half-power points North and South to check pointing accuracy. The radio galaxy 3C~123 was used as a calibrator. Its flux density was seen to vary by less than 2 \%.

One or two daily scans were carried out in the months after the campaign up to MJD 54741 (October 1st, 2008). Scans where the two receivers differ by more than 10\% were rejected. The daily fluxes from both receivers were used to calculate average flux densities and to estimate the uncertainty. A systematic error of 6\% - 7\% was estimated to account for uncertainties in the data acquisition and in the data reduction procedure. 

The Australian Telescope Compact Array (ATCA) is located at the Narrabri Observatory and consists of six 22 meter antennas\footnote{website: http://www.narrabri.atnf.csiro.au }.
ATCA data at 3 cm (HPBW $\sim$ 5'), 6 cm (HPBW $\sim$ 9'), 13 cm (HPBW $\sim$ 20') and 20 cm (HPBW $\sim$ 31') are included in Figure~\ref{fig:lc_radio}. Data were taken in the snapshot observing mode, which does not provide angular resolution. Only averages of observations from July 28 to August 4 are available. The observations start at about 14:00 UT and do not overlap with the H.E.S.S. observational window. No significant nightly flux variation was found, but the overall flux density is higher than that detected in previous observations from mid 1997 to mid 2000.

\section{MWL spectral and flux evolution}
\label{section:results}

\subsection{Flux evolution in different energy bands and correlations}

The VHE and X-ray light curves and a sample of the UV and optical light curves available from MJD 53945 to MJD 53953 are compiled in Figure~\ref{fig:multilambda}. 
All data have been binned in 10 minute bins, with the exception of the light curve in the uppermost panel, which provides the nightly averaged flux observed with the
H.E.S.S. telescopes and the average fluxes detected with the CANGAROO experiment, which correspond to a livetime of typically a few hours \citep{Sak2008}.
Only the nightly averaged three-fold triggers are shown for CANGAROO. To allow a more direct comparison, the published fluxes from CANGAROO, measured above 660 GeV, 
have been extrapolated down to the same threshold as the H.E.S.S. integrated fluxes (300 GeV). The photon indices determined with H.E.S.S. from the data sets closest in time to 
the CANGAROO measurements were used in this extrapolation. The CANGAROO data complement the H.E.S.S. data and indicate that the source was in a low state just before 
Flare 2 occurred.

A first look at the other wavelength bands reveals a high flux in the X-ray band during the night of Flare 2. Observations with {\it CHANDRA} started only after the
peak in the VHE flux and show a decreasing flare simultaneous to Flare 2. Data from {\it RXTE} and {\it SWIFT} XRT, taken over much shorter time intervals than the data from
{\it Chandra}, reveal an X-ray flux during this night that is higher than on the previous and on the following night and that is on the decrease. The optical flux is increasing during 
that night (MJD 53946), but is lower than during several of the other nights.

For the {\it SWIFT} XRT and {\it RXTE} data sets, light curves have been extracted between 0.3 and 4 keV and between 2 and 10 keV, respectively, in intervals of 10 minutes. For statistical reasons, nightly averaged spectra were used to calculate the absolute values of the de-absorbed fluxes. It is assumed that intra-night variations in the photon index are negligible for the determintation of the integrated fluxes.
The XRT and {\it RXTE} light curves can be seen to follow the same evolution. The X-ray flux is in its lowest state on the first night (MJD 53945) and shows its largest increase on the night of Flare 2, although X-ray data are only available after the peak observed in the VHE range. In the nights following Flare 2, the X-ray flux is decreasing more slowly than the VHE flux, while staying always above the low level of the first night. 

In the UV and optical bands, the flux densities have been multiplied with the effective frequencies of the corresponding bands to allow an easier comparison of the energy output at different wavelengths.  (It should be noted that the optical filter in the V-band used by {\it SWIFT} UVOT is different from the one used with the Bronberg and Watcher telescopes.) The fluxes included in Figure~\ref{fig:multilambda} have not been de-reddened.
The UV and optical flux increases during and after the night of Flare 2 and stays above its initial level in the following two or three nights, but the small amplitude of this increase 
is comparable to variations observed on other nights. 

Peak to valley flux variations of about a factor of four can be seen in the X-ray range over the selected period, whereas the VHE flux is increasing by up to two orders of magnitude. Variations in the optical range are much smaller, of the order of 30\% in the V band.

\subsubsection{VHE and X-ray flux evolution}
\label{subsubsec:vhe_vs_x}

 \begin{figure*}
     \centering
      \includegraphics[width=16 cm]{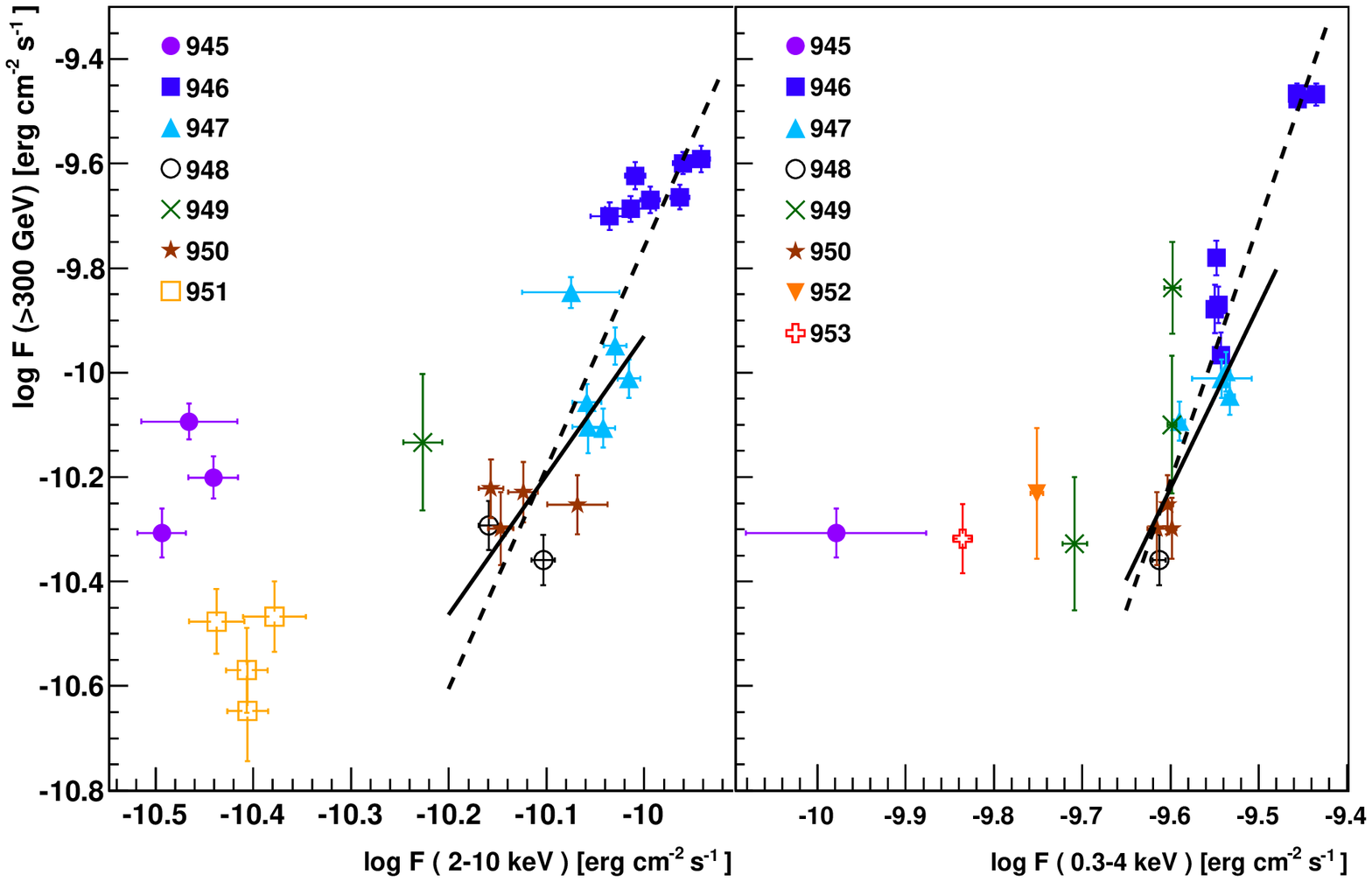}
  \caption{Correlation between the integrated VHE flux and the integrated {\it RXTE} (left) and {\it SWIFT} XRT fluxes (right). 
  The labels indicate the dates of observations (MJD - 53000).  The points represent averages over simultaneous time intervals of 10 minutes (except for the H.E.S.S./{\it RXTE} point from MJD 53949 and the H.E.S.S./XRT point from MJD 53952 that correspond to a 30 minute time window, as well as the H.E.S.S./XRT point from MJD 53953, corresponding to a 90 minute interval). The statistical uncertainties on the logarithmic fluxes are approximated with symmetric error bars.
 The two lines in each panel show power law fits to subsets of the data: the dashed lines include all ``high flux'' data, while the fits marked with solid lines include all ``high flux'' data except for the data from the night of Flare 2 (MJD 53946).} 
     \label{fig:corr_x_vhe}
    \end{figure*}

It has previously been shown that the VHE $\gamma$-ray flux correlates with almost the third power of the X-ray flux observed by {\it CHANDRA} during Flare 2  \citep{Aha2009b}. 
When examining the evolution of the X-ray and VHE flux (Fig.~\ref{fig:corr_x_vhe}) during the other nights of the campaign, one can see that, also on a longer time scale, variations in the VHE flux correlate to variations in the X-ray flux observed with {\it RXTE} and {\it SWIFT} XRT. 

When evaluating the correlation between {\it RXTE} and H.E.S.S. over the whole campaign (using a 10 to 90 minute binning, see Fig.~\ref{fig:corr_x_vhe}), a correlation coefficient\footnote{Correlation coefficients here and in the following sections correspond to the ``Pearson r''.} of 0.75 was found, corresponding to a chance probability of less than 0.001\%~\citep{Pug1966}. The comparison of H.E.S.S. and XRT data yields a correlation coefficient of 0.72 with a chance probability of less than 0.03\%. When ignoring all data points from the night of Flare 2, the remaining data still yield correlation coefficients of 0.61 ({\it RXTE}) and 0.54 (XRT) with chance probabilities of less than 0.5\% and less than 5\%, respectively.  

Power law fits to subsets of simultaneous data points are shown in Fig.~\ref{fig:corr_x_vhe}. They do not provide a good description of the data in statistical terms, hinting at a more complicated behaviour, but they show a general trend of the correlation pattern. For both data sets, data points from nights with relatively high fluxes have been fitted to illustrate this trend. Arbitrary values of log F(2-10keV)  $> -10.2$ log(erg cm$^-2$ s$^-1$) and log F(0.3-4 keV) $> -9.65$ log(erg cm$^-2$ s$^-1$) have been chosen to define the subset of data with relatively high fluxes.
The fits to the ``high flux'' points have very steep slopes, with exponents between 4 and 5 (dashed lines). When removing all points from the night of Flare 2, the data from the ``high flux'' states can be fitted with exponents between 2.5 and 3.5 (solid lines) , still presenting a steep correlation between the VHE and X-ray bands.

The power law fits indicate that the observed correlation between VHE and X-ray fluxes varies over the period of the MWL campaign. At high flux levels, the VHE flux varies as a high power of the X-ray flux variation.
The correlation becomes less steep with decreasing flux and there is no clear correlation for the nights with the lowest flux levels. This indicates a different behaviour of the emission depending on the flux level. It is also coherent with a recent joint H.E.S.S./Fermi/{\it RXTE}/ATOM campaign in 2008 \citep{Aha2009a}, where the source was found in a relatively low state and no correlation between 
the X-ray and VHE band was detected, similar to the 2003 low state. Taken together with the behaviour described here, this might indicate that a correlation between these bands occurs only if the source is in a high state. 

It should be noted that evidence of a difference in the behaviour of the emission spectrum of this source between low- and high-flux states was also found by~\cite{vhe_paper} concerning the evolution of the VHE spectral index with the VHE flux level. 

A correlation between the X-ray and VHE fluxes is naturally accounted for in the SSC framework and has been detected for example during high-flux states of Mrk 421 \citep{Kra2001, Foss2004, Tan2004, Bla2005} and Mrk 501 \citep{Cat1997, Pia1998, Dja1999, Alb2007}. An extension to more than one emission zone has been proposed to account for the observed steep correlation between the X-ray and VHE $\gamma$-ray flux during Flare 2 \citep{Aha2009b}. An application of multi-zone SSC models to Flare 2 will be demonstrated in Section~\ref{section:modelling}. Such a scenario could be applicable more generally to explain the changing behaviour of the source between high- and low-flux states, as will be discussed in Section~\ref{section:discussion}.

\subsubsection{Optical flux evolution}

As can be seen from the ROTSE data in Fig.~\ref{fig:lc_rotse}, the optical flux from the source was at a relatively high level during the 2006 MWL campaign, compared to the low state observed with ROTSE in 2003 \citep{Aha2005b}. This is also visible in the ROTSE long-term light curve (Fig. \ref{fig:lc_radio_rotse_hess}) and is confirmed by a compilation of archival data and by observations with the robotic 60 cm telescope REM in 2005 \citep{Dol2007}, indicating typical fluxes from PKS~2155-304 in the V band to lie between about 16.5 and 26 mJy. During the 2006 campaign, fluxes of roughly 27 to 33 mJy were observed by the Bronberg observatory and {\it SWIFT} UVOT. 

On intra-night time scales, no exceptional variations are observed in the optical flux during the active VHE state of the source. In the Bronberg, Watcher and ROTSE data, intra-night variations of $\approx$0.1 mag can be seen for several nights during the MWL campaign and also during the following months. The minimum time scale for these variations is of the order of 1 hour, corresponding to the width of the small peak in the Bronberg light curve around MJD 53987.85. The amplitude of the observed variability is similar to observations during the 2003 low state.

During the night of Flare 2, the Bronberg data show an increase of $\approx$0.1 mag (see Fig.~\ref{fig:lc_bronberg}). Over the following nights, the optical and UV flux rise to a local maximum that occurs probably two or three nights after the flare, judging from a comparison of the available Bronberg, Watcher and {\it SWIFT} data (see Fig.~\ref{fig:multilambda}). Several similar episodes of flux increases over a few nights are seen in the Watcher and ROTSE data and seem to be a general feature of the optical flux for this source \citep[see e.g.][]{Ryl2006}.

The correlation coefficients between the integrated H.E.S.S. and Bronberg fluxes for the different nights, taken at face value, indicate a positive correlation for MJD 53945 (0.76 for 34 points) and for MJD 53952 (0.98 for 6 points), an anti-correlation for MJD 53946 (-0.74 for 41 points), no correlation for the nights of MJD 53947 and 53951. The ubiquitous variations in the optical band
lead to random instants of correlations and anti-correlations with the VHE band. When taking into account the overall behaviour of the optical flux for all the nights where detailed optical light curves were available, there is no compelling evidence of a direct correlation of the optical band with the VHE band. 

However, on a time scale of several weeks, the average optical flux was rising, as can be seen in the Watcher and ROTSE data and by comparing the three September nights in the Bronberg data to the observations from July and August. During the Watcher observation period, the flux increased by $\approx$0.6 mag, with the fastest increase being $\approx$0.4 mag in less than 10 days, occurring after the MWL campaign, when the VHE flux was seen in a rather low state \citep{vhe_paper}. The ROTSE data, which cover a longer time span, show a fluctuation of $\approx$1.2 mag between the lowest and highest fluxes in 2006. The source was observed with REM from August 23 (MJD 53970) on and an increase to a particularly bright state on October 17 (MJD 54025) of V $=$ 12 mag was found, followed by a rapid decrease. Variations in the V band of about 0.6 mag were detected by {\it SWIFT} UVOT \citep{Fos2008}. The long term rise seen by ROTSE, Watcher, REM and Bronberg is not reflected in the VHE $\gamma$-ray data, but 
suggests a relation between high optical states and flaring VHE activity.

During the 2008 MWL campaign on PKS~2155-304, some evidence was seen of a direct correlation between the VHE and optical fluxes on a time scale of several nights. If such a correlation during the low state of the source is confirmed by future observations, it would be another piece of evidence of a difference in the emission characteristics between the high 
and low states of the source. Stronger statements on this issue require long-term MWL monitoring of the source, not only covering flaring states, but also regularly sampling low flux states.

 \subsubsection{Radio flux evolution}

   \begin{figure*}
     \centering
      \includegraphics[width=18 cm]{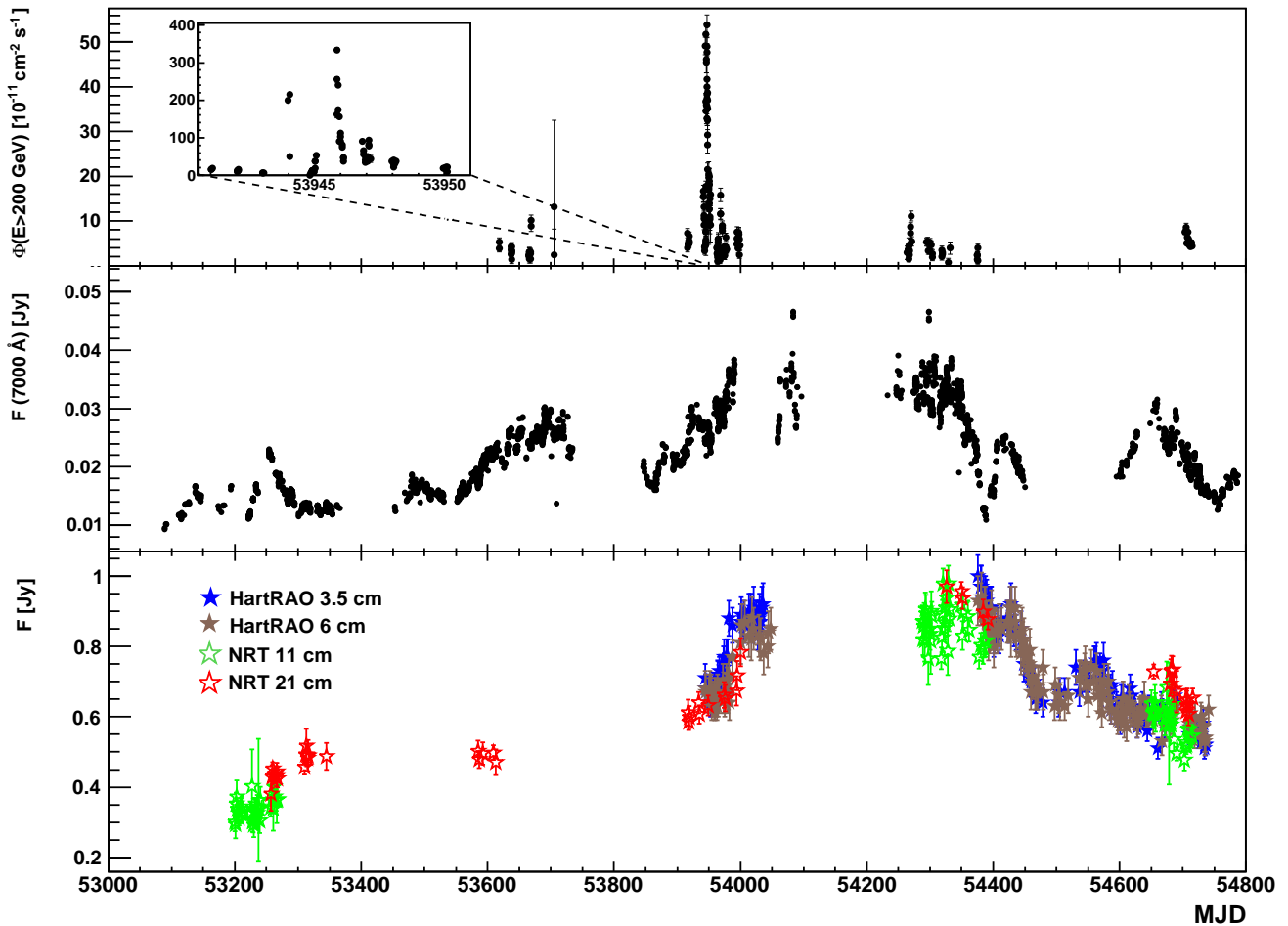}
  \caption{Long-term light curves from VHE $\gamma$-ray measurements (upper panel: H.E.S.S.),  optical data (middle panel: ROTSE) and radio data (lower panel: NRT and HartRAO). Please note that the y-axis is truncated for the long-term H.E.S.S. light curve due to the large dynamic range caused by the flares. A zoom on the period covering the two VHE flares is shown in an inset in the upper panel.} 
     \label{fig:lc_radio_rotse_hess}
    \end{figure*}

The evolution of the available radio data (Fig.~\ref{fig:lc_radio}) shows that the source was already in an active state in the radio band during the H.E.S.S. campaign and exhibited a significant increase in the flux level about one and a half months after the VHE flares occurred. No significant variation was seen between the nightly ATCA fluxes during the MWL campaign, which is confirmed by the constant flux in the HartRAO data. The average flux measured by the NRT during the campaign (0.49$\pm$0.04 Jy at 11 cm) is significantly higher than the flux measured during the 2003 campaign (0.30$\pm$0.01 Jy at 11 cm). This is consistent with the ATCA flux being higher than previous measurements found in the ATCA database. 

In Figure~\ref{fig:lc_radio_rotse_hess}, the long-term light curve in the radio band (NRT, HartRAO) is compared with the optical light curve from ROTSE and with data from H.E.S.S., taken from 2005 to 2007.
The rise in the radio flux in the months around MJD 54000 (September 2006) by roughly 50\%, observed by both the NRT and HartRAO telescopes, coincides with a comparable increase  in the optical flux during that time period. It can also be seen that both the radio and optical fluxes were at a lower level around MJD 53200  (July 2004)  and decline again around MJD 54400 (October 2007), even though in this case the relative change in the amplitude is not the same. Both bands exhibit a relatively high flux level that increases still during or shortly after the VHE high state. Although the optical and radio flux evolution shows some differences, especially around MJD 53300 and MJD 54400, which might indicate a delay between the two bands that is difficult to quantify due to the sparse data coverage, an overall correlation is clearly present. Over the whole period presented here, a correlation coefficient of 0.29 for 128 points is derived, corresponding to a chance probability of 0.1\%.

More recent data from the Effelsberg radio telescope show that the radio flux from PKS~2155-304 was continuously decreasing further in 2009 and 2010 \citep{Ang2010}.

\subsection{Spectral variability}
\label{subsection:specvar}

Spectral hardening with an increasing flux level is usually explained with the injection of highly energetic particles into the emission zone or with rapid particle acceleration. The observation of spectral hardening at both X-ray and TeV energies, observed for example during the 1997 flare of Mrk 501 \citep{Pia1998, Tav2001},  is evidence of a common origin of the variability in those bands, as expected in the SSC framework.

In the case of PKS~2155-304, spectral hardening with increasing flux had been found in the {\it RXTE} data during the low state in 2003 \citep{Aha2005b} and in the {\it RXTE} and {\it SWIFT} XRT data during the low state in 2008 \citep{Aha2009a}, but no significant spectral variability was seen in the VHE band at that time. 
On the other hand, the H.E.S.S. data taken during the four nights with the highest fluxes of the 2006 high state (MJD 53944 to MJD 53947) show a clear hardening of the VHE spectrum with flux increase.
Finally, the analysis of H.E.S.S. data from 2005 to 2007 confirms that the source does not exhibit the same spectral behaviour at high- and low-flux states \citep{vhe_paper}.

\begin{figure}
     \centering
\resizebox{\hsize}{!}{\includegraphics[width=\columnwidth]{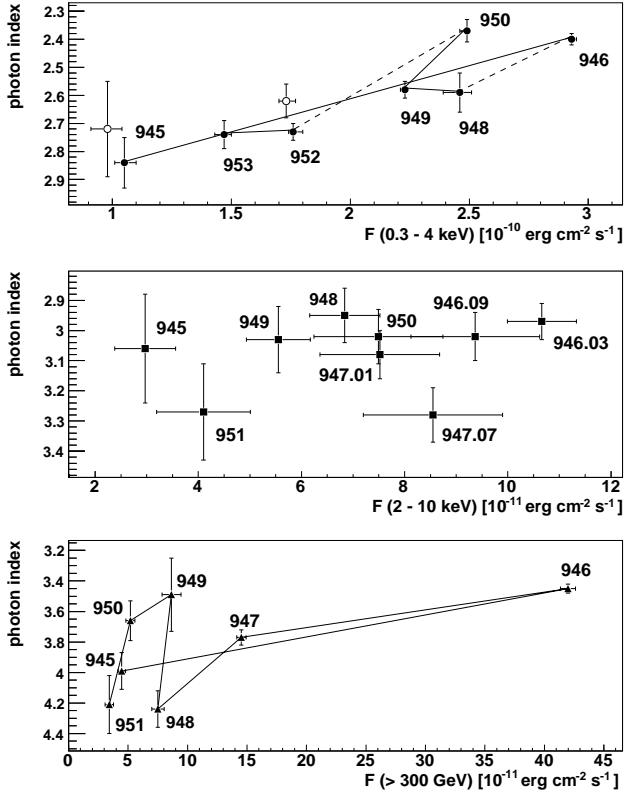}}   
  \caption{Variation of the photon index with the integrated flux. The upper panel shows the variation in the soft band of the {\it SWIFT} XRT data, the middle panel in the {\it RXTE} data, and the lower panel in the H.E.S.S. data. 
  Labels indicate the date in MJD-53000. The XRT points marked with filled circles show the photon index below the break for nights where a broken power law yields the best result and the photon index for a simple power law for the remaining 
  nights. For MJD 53945 and 53952, solutions for the XRT adopting an {\it ad hoc} break at 1 keV are shown with open circles. For more details, see Tab.~\ref{tab:xrt}. The H.E.S.S. point with the label ``951'' refers to an average over data taken from 
  MJD 53951 to 53953.}
  \label{fig:photon_indices}
    \end{figure}

The MWL spectral evolution during the 2006 campaign has so far only been studied on the short intra-night time scales of Flare 2, where spectral hardening is seen in both the X-ray and VHE band \citep{Aha2009b}. 
The spectral behaviour of nightly averaged VHE fluxes over the whole 2006 MWL campaign, in comparison with X-ray fluxes from the same period, is shown in Fig.~\ref{fig:photon_indices}.
The {\it RXTE} data show no indication for a correlation between the photon index and the hard X-ray flux (integrated between 2 and 10 keV).
Spectral hardening can be seen in the {\it SWIFT} XRT data in the soft photon flux below the break energy of $\approx$1 keV (cf. Tab.~\ref{tab:xrt} in Section~\ref{subsec:xraydata} ).  
Even though the overall change in the photon index is very modest (from 2.84 to 2.37), the correlation coefficient of 0.93 indicates a very significant effect. This correlation remains significant, albeit with a smaller coefficient of
0.86, if one imposes an {\it ad hoc} break in the spectrum at 1 keV for the data from MJD 53945 and MJD 53952 to be able to apply a broken power law for most of the nights (cf. Tab.~\ref{tab:xrt}).
Above the break energy, no significant variation in the photon index is seen in the XRT data. Thus the 2006 high state of the source seems to be characterized by spectral hardening only in the soft X-ray range. Faster cooling at higher 
energies might suppress the effect of spectral hardening in the hard X-ray band. 

During the nights of the MWL campaign, the spectral index of the nightly averaged H.E.S.S. spectra changed by $\approx$0.8. For this selection of nights, the overall correlation coefficient of 0.59 is not significant, but the spectral evolution gives a hint of a pattern more complex than a linear correlation. The H.E.S.S. data show some indication for a counter-clockwise loop in the last four data points.  On a shorter time-scale, a similar behaviour has already been seen in the VHE range  with the MAGIC telescope during a flare from Mrk~501\citep{Alb2007}. 
Features in the form of a hysteresis are generally expected in X-ray flares that are subject to synchrotron cooling \citep{Kir1998} and have been observed for example with {\it SWIFT} XRT during flares in Mrk~421 in 2006 \citep{Tra2009}, or
in PKS~2155-304, with the LAC on board the Ginga satellite in 1993 \citep{Sem1993} and with ASCA in 2000 \citep{Katao2000}. Evidence of loops have also been found recently in the high energy range (above 100 MeV) with the Fermi telescope in the case of the flat spectrum radio quasar 3C 273~\citep{Abd2010}.  

In the X-ray range, the emergence of clockwise and counter-clockwise loops is well understood. According to \citet{Kir1998}, a clockwise loop can arise when cooling is important, but the acceleration time scale is much faster than the cooling time scale, which is the case if the system is observed below the synchrotron peak. 
A counter-clockwise loop indicates that the acceleration and cooling rates are comparable, indicating that the system is observed closer to the maximum frequency. 
In this case, the gradual acceleration of particles leads to increased emission first at lower and then at higher energies. In the {\it SWIFT} XRT data set presented here, no conclusion can be drawn on the existence of such features, due to the low statistics and the missing spectral information for MJD 53947.

In the SSC framework, where the (VHE) $\gamma$-ray emission is strongly connected with the synchrotron emission, loops in the VHE data could reflect related features in the X-ray range.
If such patterns, as suggested by the present data, were confirmed in the VHE range by future instruments with higher sensitivity, such as the Cherenkov Telescope Array \citep{cta_cdr}, this could provide important additional constraints on the validity of the SSC scenario. 

\section{Modelling of the MWL data on different time scales}
\label{section:modelling}

The MWL coverage of the source over several nights makes it possible to investigate the spectral energy distribution (SED) and its evolution on two different time scales. 
Night-to-night variations can be evaluated with stationary models for non-flaring nights. Conversely, the night of Flare 2 allows the study of the evolution of the SED on intra-night time scales. 
Interpreting the variability observed during this night requires time dependent modelling. 

In Section~\ref{subsec:stationary}, a conventional stationary SSC model is used to describe the nightly averaged SEDs, while the flux evolution during Flare 2 is illustrated using different examples for
time-dependent SSC scenarios in Section~\ref{subsec:dynamic}. Basic considerations on the requirements of a time-dependent scenario for the night of Flare 2 \citep{Aha2009b} challenge the common 
scenarios for VHE blazar emission and show that standard one-zone SSC models cannot account for the detected SED evolution. 
Here, three different approaches to time-dependent SSC modelling with increasing complexity are discussed, which are representative of the most common options available in the literature beyond one-zone models. The first scenario is a basic description of an extended VHE-emitting source, well adapted to fast variability processes, focusing on high-frequency emission. The second one considers a two-component jet to reproduce the SED and its evolution. The third one proposes a stratified jet with time-dependent particle injection and acceleration. The first two scenarios assume that the initial particle acceleration is due to first order Fermi processes, while the third one favors a second order Fermi mechanism. 

The intention of the time-dependent modelling of Flare 2 presented in this section is to illustrate how different SSC approaches can arrive at a good characterization of the dominant components in the flux variation and what constraints the data provide in general on physical source parameters. These models present three possible ways of describing the data set and do
by no means exclude alternative descriptions. A description of the very rapid variability observed during Flare 1 and Flare 2 \citep{vhe_paper}, on time scales down to a few minutes, is beyond the scope of the models presented here, which focus on the general trend of the flux evolution during the night of Flare 2.

For all the models discussed in this section, absorption of the VHE $\gamma-$rays by the Extragalactic Background Light (EBL) is treated following \citet{Aha2009a}.

\subsection{Stationary modelling : one-zone SSC model}
\label{subsec:stationary}

The average SEDs for the different nights of the MWL campaign, except for the night of Flare 2, can be described with a stationary homogeneous one-zone SSC model. The X-ray and VHE bumps are interpreted respectively as the synchrotron and IC emission from a single population of relativistic electrons in a spherical plasma blob inside the jet. The host galaxy and the extended jet contribute to the SED at lower frequencies. The blob is characterized by its radius $r_\mathrm{b}$, Doppler factor $\delta_\mathrm{b}$ and  tangled magnetic field $B$  \citep[see][]{Kat2001}. The electron energy distribution is assumed to follow a broken power law with indices $n_1$, $n_2$, minimum and maximum energies corresponding to Lorentz factors $ \gamma_{\rm min}$ and $ \gamma_{\rm max}$  and a break energy corresponding to $\gamma_{\rm br}$. The parameter $K_1$ determines the normalization of the broken power law at a Lorentz factor of 1. This type of SSC model has typically eight main free parameters and thus requires high quality data over a large spectral range to be well constrained. 

\begin{figure}[h!]
      \resizebox{\hsize}{!}{\includegraphics[width=\columnwidth]{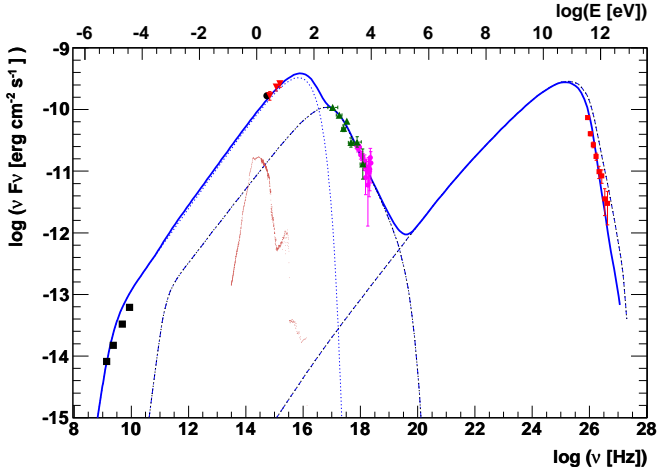}}
  \caption{SED for a night with relatively low VHE and X-ray flux (MJD 53945) with a stationary SSC model described in the text.  
  Data taken during this night are shown from H.E.S.S. (red squares), {\it RXTE} (magenta circles), {\it SWIFT} XRT (green triangles), {\it SWIFT} UVOT (red triangles) and Bronberg (black circle). The black squares are the average ATCA points from the period of the MWL campaign.
  The model is presented as a solid line, giving the overall SED,  including absorption by the EBL. The dashed line in the VHE range is the de-absorbed spectrum. The H.E.S.S. data points represent the measured (i.e. absorbed) flux. The extended jet emission (dotted line) and host galaxy contribution (brown data points) are also shown at low frequencies. Dash-dotted and dashed lines indicate the synchrotron and SSC contributions from the blob, respectively.}
      \label{fig:sed_stat_945}
\end{figure}

The available data on the nights of the MWL campaign do not provide enough constraints on the SEDs to narrow down the parameter space even of  such a stationary model.
Therefore, a set of parameters derived from the 2003 H.E.S.S. campaign (cf. ``Model 2'' in~\cite{Aha2005b}) was used as a starting point for the modelling of the 2006 data of the source, in order to compare the results for 2003 and 2006. 
These parameters have been adapted to the nightly averages of the 2006 campaign by applying only minimal  changes to the initial values. A set of parameters that reproduces the average SED of the relatively low state on MJD 53945 is presented in Table~\ref{tab:stationary}. Even though the VHE $\gamma$-ray flux during the whole 2006 observation period is higher, the SED of MJD 53945 (Fig.~\ref{fig:sed_stat_945}) is not very different from the one recorded in 2003. For the same Doppler factor and magnetic field strength, a slight increase of the size and density of the emitting blob and the spectral index $n_2$ provides a good match of the SED observed during MJD 53945. 

\begin{table}[h!] 
  \centering
  \begin{tabular}{p{0.5\linewidth} c}
    \hline
   \multicolumn{2}{ c }{\mbox{plasma blob}}  \\
    \hline
    $\delta_\mathrm{b}$  & 25  \\
    $B$ [{\rm G}]  & 0.25 \\
    $r_\mathrm{b}$ [{\rm cm}]  & $3.2 \times 10^{15}$  \\
    $K_1$ [{\rm cm}$^{-3}$] &  $6.0 \times 10^3$   \\
    $n_1$ & 1.8  \\
    $n_2$ & 5.0 \\
    $\gamma_\mathrm{min}$  & 1.0   \\
    $\gamma_\mathrm{br}$  & $6.0 \times 10^4$  \\
    $\gamma_\mathrm{max}$  & $9.0 \times 10^5$   \\
    $N_\mathrm{tot} \sim K_1 {r_\mathrm{b}}^3$  & $3.21 \times 10^{49}$ \\
    \hline
    \multicolumn{2}{ c }{\mbox{extended jet}}  \\
    \hline
    $\delta_\mathrm{jet}$  & 2.0  \\
    $B_{\mathrm{jet}}$ [\rm{G}]  & 0.06 \\
    $r_\mathrm{jet}$ [{\rm cm}]  & $3.0 \times 10^{16}$  \\
    $l_\mathrm{jet}$ [{\rm pc}] & 55 \\
    $K_{\mathrm{jet}}$ [{\rm cm}$^{-3}$] &  400    \\
    $n_\mathrm{jet}$ & 1.7  \\
    $\gamma_\mathrm{max}$  & $2.0 \times 10^5$   \\
    \hline   
  \end{tabular}
  \caption{Parameters used for stationary SSC modelling of the nightly average spectrum of MJD 53945. A description of the parameters is given in the text.}
  \label{tab:stationary}
\end{table}

The difference between $n_1$ and $n_2$ is larger than what would be expected for a pure synchrotron cooling break in the electron spectrum; the latter does not yield a satisfactory presentation of the data. The broken power law here is a simple parameterization of the electron spectrum, which may be interpreted as a sum of several effects that are not explicitly modelled, e.g. synchrotron and IC cooling, adiabatic expansion, energy-dependent acceleration and escape times, inhomogeneous particle distribution, etc.

Given the absence of direct correlations between the high-energy (VHE and X-ray) and low-energy (optical and radio) bands in the data, the radio component, and the main part of the optical emission is ascribed to a different origin than the 
high-energy bumps. It can be modelled as the emission of the extended jet \cite[see][]{Aha2005b}, with an additional contamination by the host galaxy.  A relatively low electron density and magnetic field in the extended jet results in a synchrotron emission peak from the radio to the UV range that is accompanied only by a negligible IC component at X-ray energies. For the extended jet shown in Fig.~\ref{fig:sed_stat_945}, a jet opening angle and angle to the line of sight of 2 degrees have been assumed. The other main parameters, such as the radius $r_\mathrm{jet}$, length $l_\mathrm{jet}$, the initial magnetic field  $B_{\mathrm{jet}}$ and electron distribution, are listed in Table~\ref{tab:stationary}. A more detailed description of all the parameters in this inhomogeneous jet model can be found in~\cite{Kat2001}.

The contribution of the host galaxy, as shown in Fig.~\ref{fig:sed_stat_945}, has been deduced from the magnitudes given in \cite{Kot1998}, assuming a low redshift solar metallicity elliptical galaxy of age equal to 13 Gyrs (R-H=2.4), corresponding to a mass of 5$\times$10$^{11}$ solar masses \citep{Fio1997}. The galaxy is clearly not contributing much in the optical, even at a state of low activity.

One can find good solutions for all the remaining nights (MJD 53947 to 53950) after Flare 2, with the same bulk Doppler factor and magnetic field and a similar electron distribution, while only changing the radius of the blob (by less than $\approx$ 40\%), the normalization of the electron spectrum (by up to a factor of 2.4) and $n_2$ (change in index of up to 0.2). The indicator for the electron number $N_{\rm tot}$ can be kept constant during these adjustments. Such variations could arise if the shock front passed regions of different size and electron density, or for contraction and expansion of the emitting blob.  

It should be stressed that the solutions found here are not unique, due to the insufficient information from the MWL observations. Recent data from the Fermi satellite have shown the importance of covering the high-energy $\gamma$-ray range below VHE energies in this respect.

\subsection{Time-dependent modelling : Two-zone and stratified jet SSC models}
\label{subsec:dynamic}

\subsubsection{Model 1 : Light crossing time effects} 

A relatively simple approach is given by a model based on \cite{Chi1999}. A short description can be found in Appendix~\ref{subsec:app1}. This model has already been successfully applied to Flare 1 \citep{Kat2008}, but the X-ray light curve measured by {\it CHANDRA} during Flare 2 adds  constraints which were not available for the analysis of Flare 1. 
The realization of Model 1 that is in best agreement with the data of MJD 53946 is shown in Figure~\ref{fig:model_kk} and a list of the corresponding parameters is given in Table~\ref{tab:dynamic1} in 
Appendix~\ref{subsec:app1}. 

  \begin{figure*}[h!]
     \centering
      \includegraphics[width=17cm]{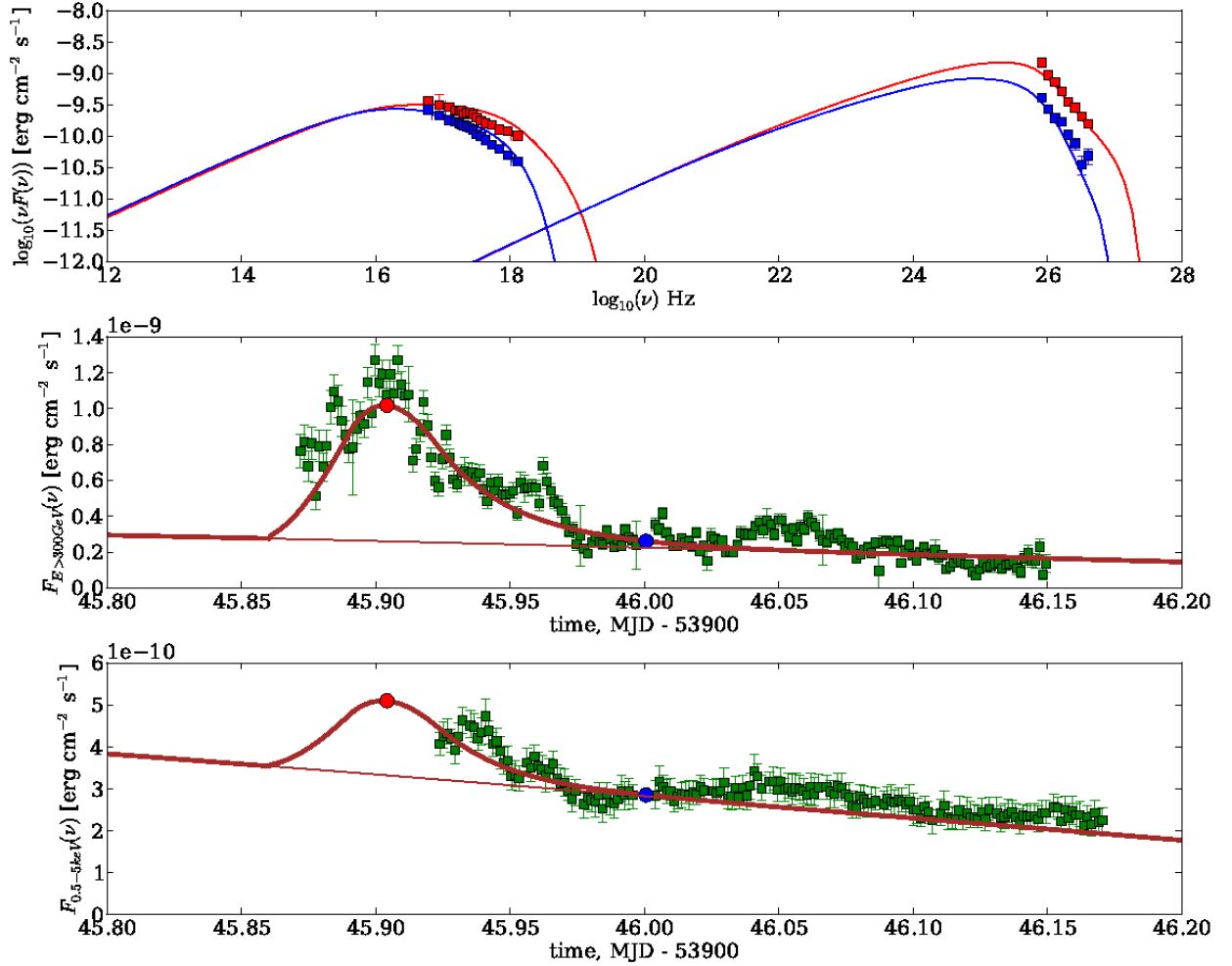}
  \caption{Time dependent SSC model (Model 1) for the night of MJD 53946. 
   The upper panel shows the modelled SED at the maximum flux level and at a low level. Snapshots of the {\it CHANDRA} spectrum and the H.E.S.S. spectrum are shown for a high state (red squares) and for a low state (blue squares). The two flux states that were chosen for the modelled SEDs correspond to the red and blue circles in the middle and bottom panel, where the H.E.S.S. and {\it CHANDRA} light curves are shown together
    with the predicted time evolution of the model. The thin line in the middle and bottom panel indicates the long-term component that dominates the X-ray flux. The two spectra from {\it CHANDRA} correspond to the beginning and 
    the plateau of the {\it CHANDRA} lightcurve, as  described by~\cite{Aha2009b}; the H.E.S.S. spectra correspond to the data acquisition periods closest in time to the modelled flux states.}
          \label{fig:model_kk}
    \end{figure*}

Two independent sources were necessary to explain the activity at high energies observed during that night. The first source has an extension of $4 \times 10^{16}$ cm and an electron density of $1.6 \times 10^4$ cm$^{-3}$ and leads to a long-term component with a time scale below two days that is visible in the X-ray flux and is consistent with the observations by {\it RXTE} and {\it SWIFT} XRT on the previous and following nights. This source dominates the synchrotron emission due to its large size, but does not contribute significantly to the IC emission due to its relatively low electron density. 
The second source is much smaller ($2.35 \times 10^{15}$ cm)  and denser by a factor of 100, with a larger $\delta_\mathrm{b}$ than for the first source. This component dominates the IC emission, while contributing little to the synchrotron flux. It is responsible for the rapid variability of the order of an hour that characterizes the main VHE flare. 
The contributions of the two components to the integrated X-ray and VHE flux are shown in the middle and lower panel of Figure~\ref{fig:model_kk}. 

In this model, the adjustment of few main parameters --- namely the extension of the emission region, magnetic field and density of the injected electrons and the bulk Doppler factor of each source --- provides a satisfactory description of the main components of the observed high-energy data during the two VHE flares. The parameters found for Flare 2 are comparable to those for Flare 1, with identical $\delta_\mathrm{b}$. Additional small and dense sources can be added to account for the structure visible on smaller time scales, as has been shown for Flare 1 by \cite{Kat2008}.
The long term component found in Model 1 does not extend over the whole duration of the MWL campaign; it is constrained by X-ray data from the night before and after the night of Flare 2. Additional large emission zones would need to be added to reproduce the behaviour of the X-ray flux detected by {\it RXTE} and {\it SWIFT} over several nights. 
The optical and radio emission is ascribed to an additional component, e.g. the extended jet, which is not included in this model. Thus the low frequency points only serve as upper limits for the SED derived with Model 1.

\subsubsection{Model 2 : Blob-in-jet scenario} 

A more refined approach proposed by \citet{Kat2003} is aimed at providing a complete description of the SEDs of blazars and of their variability from radio to VHE energies. A short description can be found in Appendix~\ref{subsec:app2}. A first application of this model to Flare 2 has been presented by~\citet{Len2008} (see also \citet{Len2009a}). 

  \begin{figure*}[h!]
     \centering
      \includegraphics[width=9cm]{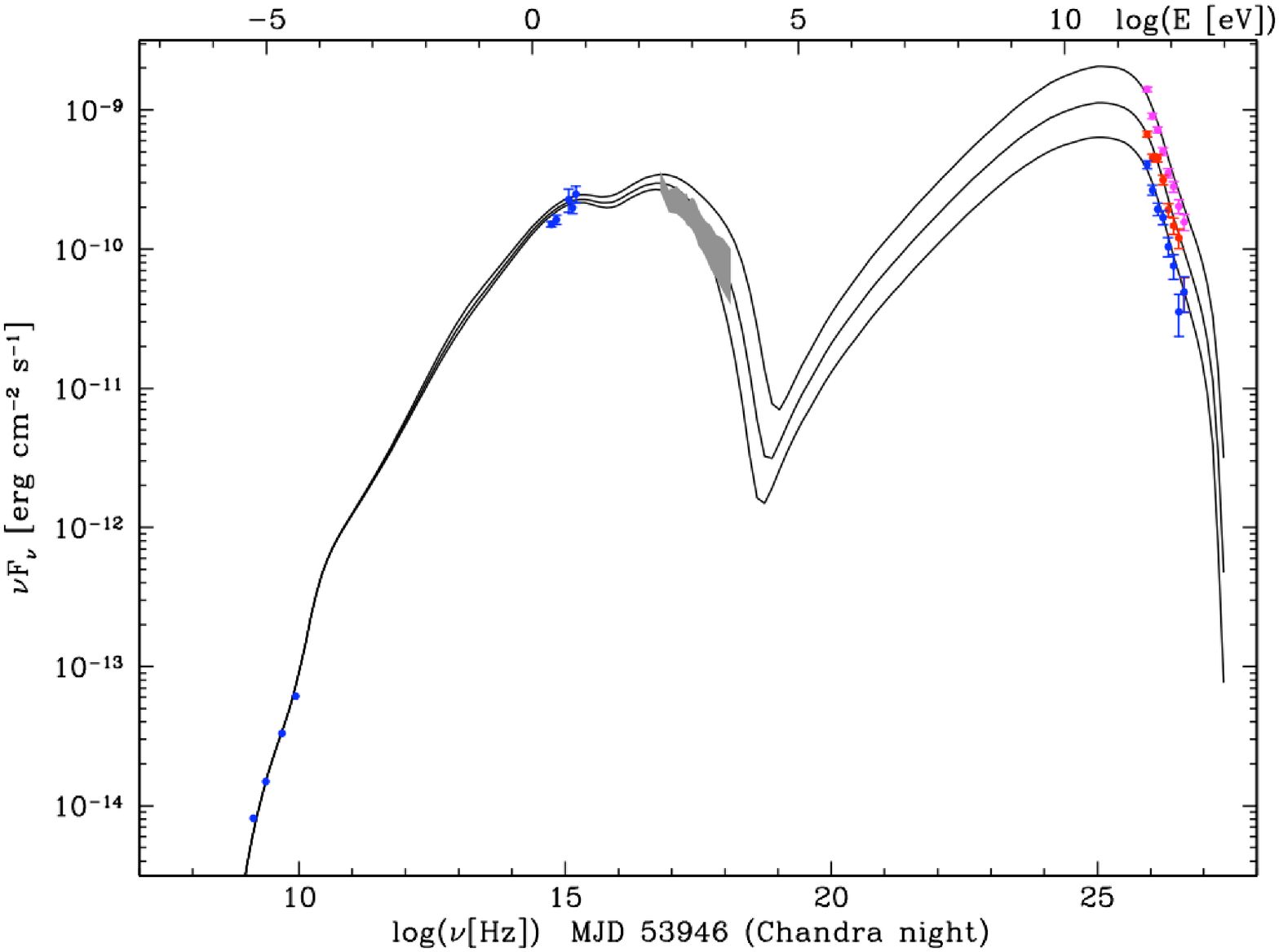}
      \includegraphics[width=9cm]{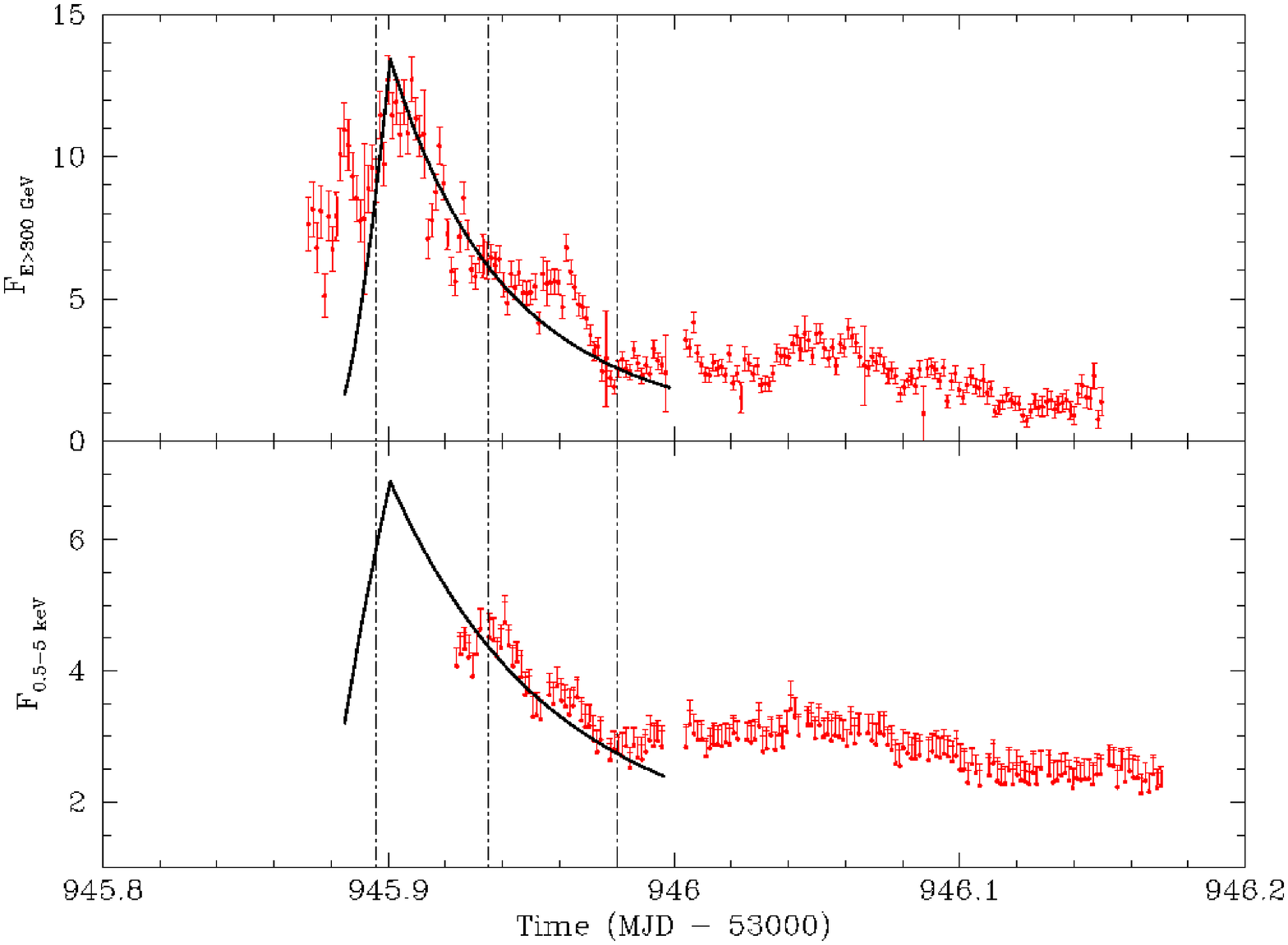}
  \caption{Time dependent SSC model (Model 2) for the night of MJD 53946.
       Left panel: H.E.S.S. spectra at three different states of the source are shown. The varying X-ray spectrum seen by the {\it CHANDRA} 
       telescope is indicated by a dark grey band. Snapshots of the SED resulting from the model for the three different states are presented by solid lines.  
       Optical and UV data from {\it SWIFT} UVOT and radio data from ATCA are shown as green points.
       Right panel: The light curves in the VHE band (H.E.S.S., upper panel) and in the X-ray band ({\it CHANDRA}, lower panel) are shown in units of 10$^{-10}$ erg cm${-2}$ s$^{-1}$, together with the predicted time evolution of the model. 
       The times corresponding to the three H.E.S.S. spectra in the left panel are marked here by three vertical dash-dotted lines.}
          \label{fig:model_jp}
    \end{figure*}

The application of the blob-in-jet model to Flare 2 is shown in Figure~\ref{fig:model_jp}, with the corresponding parameters in Table~\ref{tab:dynamic2} in Appendix~\ref{subsec:app2}.
The SED shows the spectra derived from three different H.E.S.S. data acquisition periods during that night, each of a duration of about 28 minutes, together with 
the area corresponding to the varying spectrum observed with {\it CHANDRA}. The model describes well the X-ray and VHE spectra at different flux levels during the night. 
It also reproduces well the main flux evolution at X-ray and VHE energies, as can be seen from the comparison between model and observed light curves. The observation with {\it CHANDRA} began later than the H.E.S.S. observation that night, and the rising part of the flare is not covered in the X-ray data, so the highest flux state is not constrained in the X-ray band. 

Contrary to Model 1, the two high-energy emission regions (jet and blob) of Model 2 are not fully independent. Here it is assumed that they have the same bulk Doppler factor of 50. 
The blob is denser than the jet by three orders of magnitude, with higher magnetic field, higher maximum particle
energy and smaller initial radial extension. 
The ratio of the energy density in the magnetic field and in the radiation field in the initial emission region is very large for the jet ($10^{3}$), which is the principal synchrotron source, and close to unity for the blob, which is mainly responsible for the SSC emission. 
However, the initial extension of the blob, its initial electron density and magnetic field are close to the values of the small and dense ``source 2'' in Model 1. The main differences between the two approaches lie in the global AGN scenario and in the resulting values for the bulk Doppler factors, which are larger in Model 2. This stems mainly from the necessity to compensate for adiabatic losses in Model 2, which are not taken into account in the simpler Model 1. Additional plasma blobs may be introduced to account for a finer temporal structure of the light curves. 

The radio and optical emission is reproduced by the addition of a large-scale jet with the same parameters as in the stationary model (cf. Table~\ref{tab:stationary}), except for a small decrease in the magnetic field to $B=$0.04 G. The optical emission is dominated by emission from the large-scale jet, but has also contributions from the plasma blob and inner jet. The rapid variation in the emission from the blob, which explains the X-ray and 
VHE flare, leads only to a very slight variation in the optical and UV range, smaller than the size of the statistical error bars for the nightly averaged flux.

In the optical data, variability is seen on different time scales and might be due to more than one emission region. The variability time scales can be used to constrain the size of the potential emission regions as $R \le  c \:  t_{\rm var} \:  \delta / (1 +  z)$.
For the intra-night variability seen in the Bronberg data on the hour scale, this would lead to a ratio of source extension and Doppler factor of roughly $ R \: \delta^{-1} \le 10^{14}$ cm.
The derived $\delta_{\rm blob}$ and $R_{\rm blob}$ have a ratio of the same order of magnitude, so that the plasma blob might in this model be identified as the emission zone responsible
for the intra-night variations in the optical band. A direct correlation with the emission at higher energies, which is not observed, might be washed out by the summation of several components. It is also possible 
that another component, e.g. emission from the accretion disc, is responsible for the rapidly varying optical flux.

The observed optical variability on the time scale of several days (in the Watcher and ROTSE data) and the long-term variation seen in the ROTSE flux might come from a different emission region. 
For this component,  emission regions of the size of the inner jet or large-scale jet are not excluded. The same regions might also account for the radio emission, which has been shown to be correlated with the optical flux 
on long time scales.

\subsubsection{Model 3 : Stratified jet scenario} 

The third model considered here assumes a time-dependent stratified jet. Contrary to the previous models, this model has only one radiative component, the jet itself, the global SED resulting from the integration, along the jet, of the particle flow emission. A short description can be found in Appendix~\ref{subsec:app3}. This model has already been applied to describe the H.E.S.S. light curve of Flare 1 \citep{Bou2008}. 

  \begin{figure*}[h!]
     \centering
      \includegraphics[width=9cm]{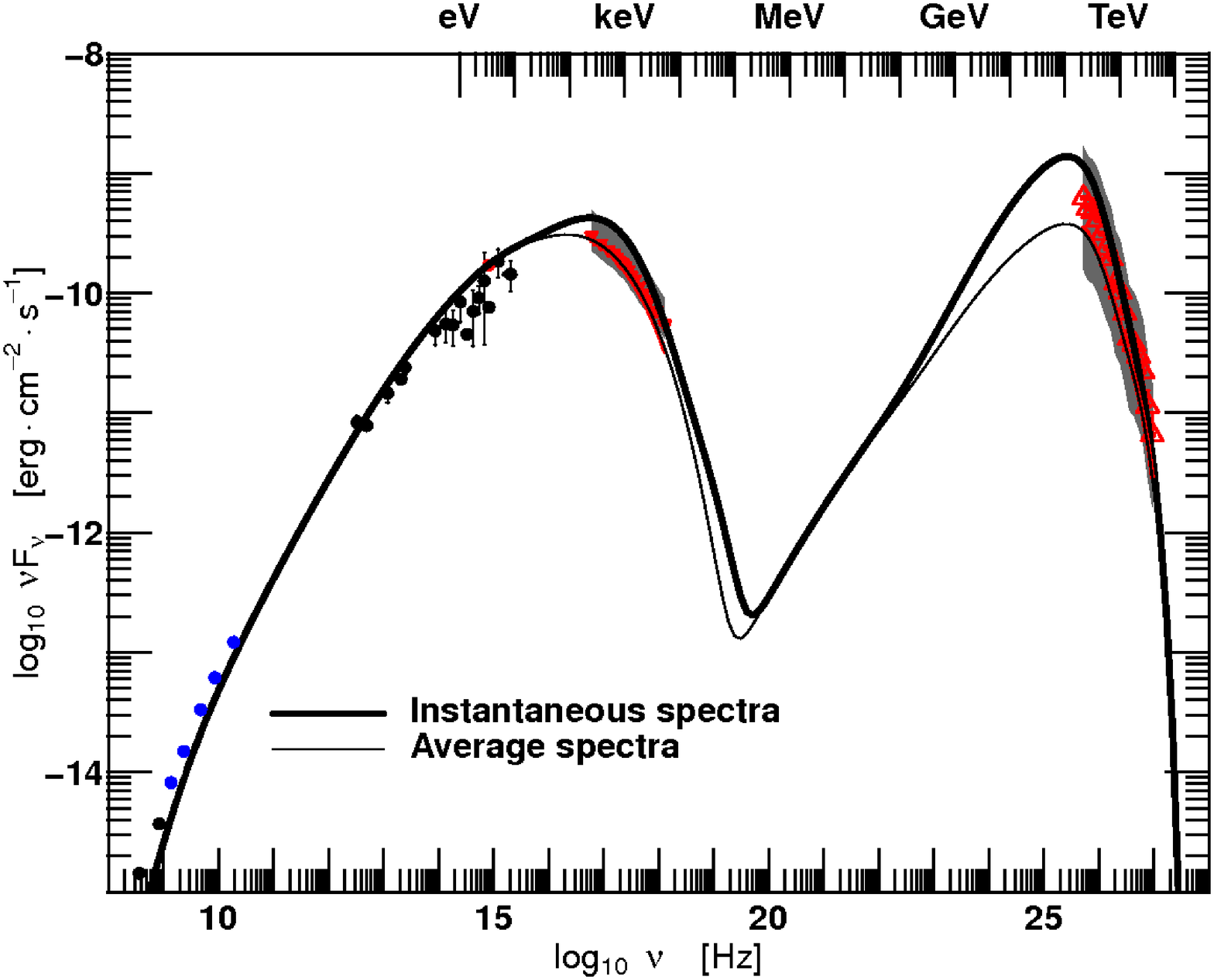}
      \includegraphics[width=9cm]{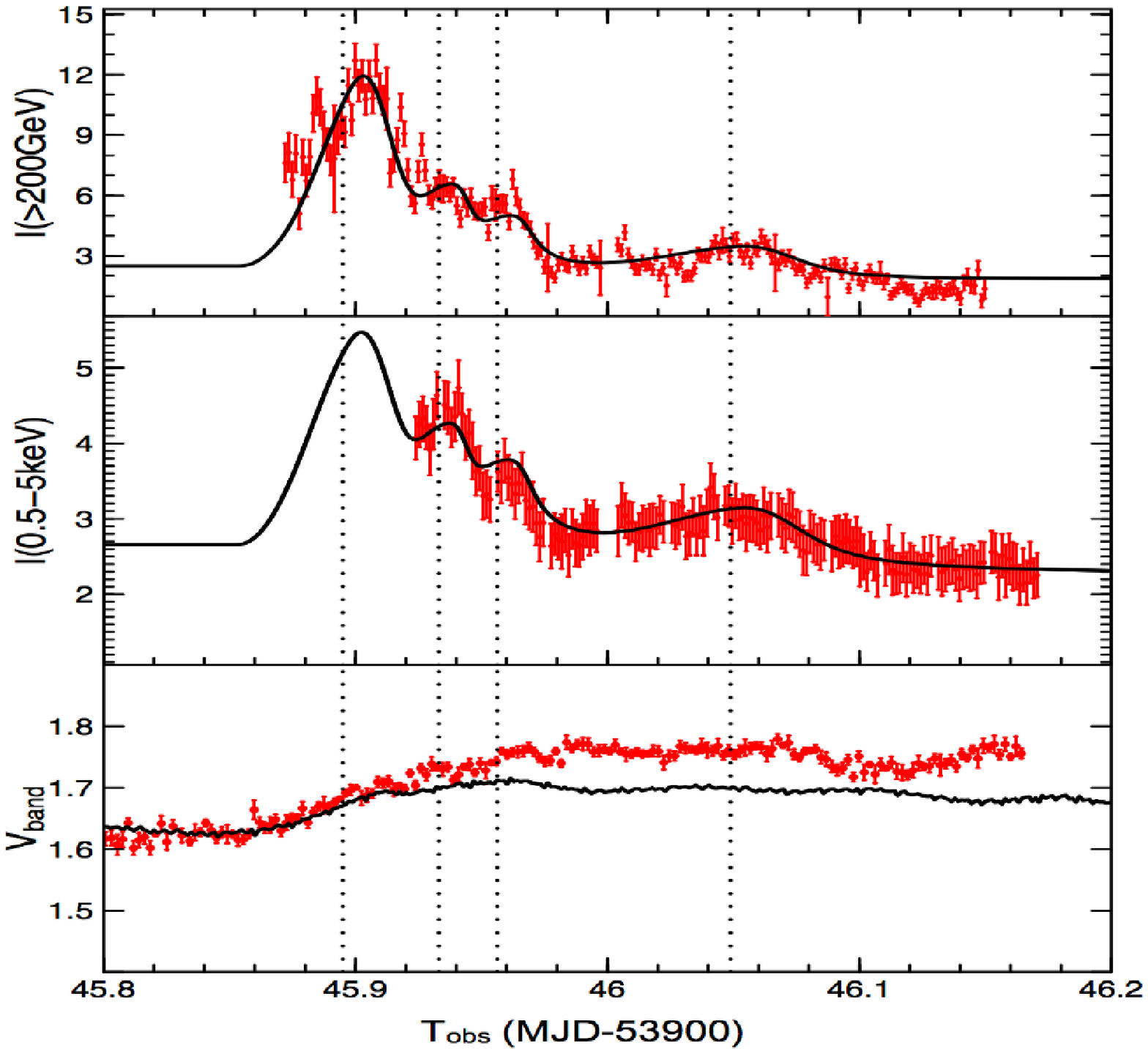}
  \caption{Time dependent SSC model (Model 3) for the night of MJD 53946. 
Left panel: SED of PKS\,2155$-$304, including the band of the varying {\it CHANDRA} spectrum and the band of the varying H.E.S.S. spectrum.
   Nightly averaged spectra for {\it CHANDRA} and H.E.S.S. are indicated by red points. The optical flux detected by the Bronberg Observatory for this night is added as a red point; data from ATCA are presented as blue points. 
   The thick line shows a snapshot of the modelled SED at a given state of the source, while the thin line shows the modelled spectra averaged over the H.E.S.S. observation period. Black data points indicate archival (non-simultaneous) data.
  Right panel: The points represent the H.E.S.S. (upper panel), {\it CHANDRA} (middle panel), and Bronberg (lower panel) light curves in units of 10$^{-10}$ erg cm$^{-2}$ s$^{-1}$. The solid lines are the light curves computed by the model in these three energy ranges. }
          \label{fig:model_timothe}
    \end{figure*}

Fig.~\ref{fig:model_timothe} shows the application of Model 3 to the data from Flare 2, with the principal model parameters summarized in Table~\ref{tab:dynamic3} in Appendix~\ref{subsec:app3}. The model reproduces very well the SED of PKS~2155-304, as well as the simultaneous light curves in the X-ray and VHE range. It reproduces more accurately the variability in X-ray and VHE light curves than the previous models, due to the freedom in rapidly varying the flux of particles from the injection and acceleration process, and it provides a full MWL prediction of the emission of the blazar. 

In the application presented here, the flux of injected particles is kept roughly constant at the base of the jet, whereas the acceleration term varies. The flux of particles in the region where most of the VHE emission comes from is one order of magnitude higher than the initial particle flux. This reveals that intense pair production has occurred between the base of the jet and the emission region, producing a strong amplification of initial perturbations in the injection parameters, and hence the strong and fast variability observed at VHE. In the framework of Model 3, pair production appears to be a likely mechanism to interpret fast variability of PKS~2155-304. In situ re-acceleration of the electron-positron plasma and the additional emission from a first generation of pair-produced leptons in Model 3 provide the observed level of luminosity during Flare 2 with a smaller Doppler factor than in Model 2. 

The model partially reproduces the optical light curve over the night, although it does not fully match the amplitude of the observed flux variation. A smooth increase in the optical band is obtained as observed, in response to the rapid variation of the electron-positron flux. However it lasts for a shorter time interval than the one observed. This increase is due to the large size of the optical emitting region that plays the role of a low pass filter. The optical luminosity integrates the recent past history of the jet. No direct correlation between the optical and the high-energy flux is expected from the model, instead the optical flux responds to high-energy events in the source with a certain delay. Realistically, the optical component should contain contributions from earlier events that occurred before the onset of the observations during MJD 53946. Given a complete knowledge of the past variations of the high-energy flux, the inclusion of a longer 
history of high-energy events in this model might, at least in theory, lead to an exact match with the observed optical light curve.

\section{Discussion} 
\label{section:discussion}

\subsection{Modelling VHE flares and high-flux states}

Given the current understanding of the source PKS~2155-304, all three multi-zone SSC models presented in Section~\ref{subsec:dynamic} arrive at a good description of the VHE and X-ray emission during the night of Flare 2. This illustrates the present success of SSC models for blazars, but also their inherent degeneracy. The model parameters are not unique solutions, but some general conclusions can still be reached.
In all three scenarios, the dominant X-ray emitting zone appears larger than the VHE emitting zone. The X-ray emission can be attributed mostly to the jet, while the VHE emission comes from a denser emission region, identified as a blob in the jet or its inner core. The flares visible in the X-ray and VHE bands are ascribed to the dense emission region, with a size in the range of 10$^{15}$ to 10$^{16}$ cm and a magnetic 
field of a few 0.01 G. High Doppler factors ($\gtrsim$30) are always needed and the high-energy tail of the particle distribution has to reach individual Lorentz factors of the order of 10$^{6}$.

A multi-zone scenario, similar to the ones described in Section~\ref{section:modelling}, but on a longer time scale, might also account for the different characteristics observed during high and low states of the source. The synchrotron emission would be generally dominated by a not very dense emission region that does not contribute significantly to the SSC component. An additional emission region with a denser and more energetic electron population would dominate the VHE emission while contributing little to the X-ray emission. This second zone would account for the steep correlation and spectral hardening during high states and flares. In a low state of the source, the small contribution from the second zone to the X-ray emission might become undetectable against the dominant emission from the first zone. This would naturally explain the uncorrelated behaviour or less steep correlations at lower flux states and the emergence of correlations during higher flux states.

The three models presented in this work account for a large Compton dominance in the emission, yielding a VHE flux variation in Flare 2 that by far supersedes the observed variation in the correlated X-ray flare. Clearly, an extreme situation could arise where the X-ray flux variation becomes too small to be detected against the emission from the larger zone and the VHE flare seems to be without any X-ray counterpart. Time-dependent multi-zone SSC scenarios can thus account for the ``orphan flares'' detected in 1ES1959+650 \citep{Kra2004} and Mrk 421 \citep{Bla2005}, as previously suggested by \cite{Kra2004} and \cite{Len2009b}. 

Complementary particle injection events can be added to Models 1 and 2 to reproduce more closely the structure of the light curves, which include at least one secondary peak during Flare 2. 
However, a description of the fastest observed variability during Flares 1 and 2 has not been attempted here. Variability of a few minutes puts strong constraints on the size of the emission region and requires a very large bulk Lorentz factor of the emitting electrons, which is also needed to allow the escape of VHE photons from the emission region \citep{Aha2007, Beg2008}. Ascribing some of the VHE emission to external inverse Compton processes might help to loosen some of these constraints \citep{Beg2008}.
Following the ``needle-in-jet'' scenario \citep{Ghi2008} or the ``minijet'' scenarios \citep{Gia2009, Nal2011}, very small emission regions inside the jet, much smaller than the plasma blobs discussed in this work, could be responsible for very rapid variability. Such emission zones might be caused by magneto-centrifugal acceleration of electron beams or magnetic reconnection in a Poynting flux dominated jet and would be moving rapidly towards the observer. An alternative explanation is given by \cite{Ner2008}, who assume a radiatively inefficient accretion flow and a relatively small mass of the central black hole to account for rapid variability.

\subsection{Low frequency emission and long-term evolution}

Additional constraints from MWL data should help to resolve the degeneracy of the models at high energies. 
However, despite their success at high energies, none of the above scenarios is able to provide a detailed description of the light curves at low frequencies. Qualitatively, Models 2 and 3 propose to interpret the optical emission as a low energy counterpart of the activity at high energies, with a dominant contribution from the extended synchrotron jet in Model 2, and integration of the recent past VHE events in Model 3. However, they do not reproduce quantitatively the short-term optical light curve during Flare 2 or the long-term optical and radio evolution. This emphasizes the current difficulty to relate the very fast and sporadic events seen at VHE in blazars to long-term properties traditionally explored in the radio and optical ranges. Various clues show that they are partially linked, but how and to which extent remains unclear. 

In this regard, the radio and optical light curves gathered over years during and after the 2006 MWL campaign are quite interesting. The radio data might indicate a similar situation to the one seen in 2008 in the radio-galaxy M87 \citep{Acc2009}. In both cases, a rapid variation at the highest energies occurred during or shortly before an increase in the radio flux over several months. From the available data it seems that the radio flux from PKS~2155-304 was at a relatively high, but stable level when the VHE flares occurred, and started to increase about one month afterwards. 

In the two-zone scenarios of Models 1 and 2, part of the radio and optical emission could come from the same lepton population that causes the VHE $\gamma$-ray and X-ray flares. The blob could be initially opaque to radio waves due to synchrotron self-absorption, as has been observed in Mrk421~\citep{Cha2006}. The expansion of the blob permits the emission of radio waves with a certain delay. 
The longer cooling times at lower energies lead to a longer time scale for radio and optical emission compared to the very rapid high-energy flares. 
Moreover, the variability is diluted in the radio light curve by emission from the inner and large-scale jets, as well as from the extended radio lobes. Diffusion of 
energetic particles from the blob into an external component could lead to a delayed response to the VHE flares in the low-energy bands. 

In the stratified jet of Model 3, the emission observed at lower energies represents the sum of different high-energy events, which are delayed and spread out in time due to the cooling, propagation and re-acceleration of leptons in the jet. As in the previous case, the plasma would be optically thick in the beginning and become transparent to radio synchrotron emission with adiabatic cooling. 
In this way, high activity at high energies and on short time scales directly contributes to an averaged, long-term signature at low energies, while in the X-ray and VHE $\gamma$-ray band, variations in the injected flux close to the base of the jet become visible as individual flares. 

For the observations of the VHE flare from M87 in 2008 no simultaneous optical data have been published, thus it is not known if the optical flux showed a behaviour similar to the radio flux. However, in the case of PKS~2155-304, the Watcher and ROTSE data show the average optical flux to be in a high state and on a long-term increase when the VHE flares occurred. Variations on the scale of a few nights and on intra-night scales are superimposed on this increase. Similar to the radio flux, the optical flux could represent a smoothed response to previous high-energy activity. 

The simulation of a much longer history of particle injection in the jet of Model 3, which is beyond the scope of the work presented here, would be needed to modify the optical and radio light curves and might account for the overall evolution in these wavebands. The sparse sample of VHE data does not permit to determine if the long-term increase in the optical range, already seen before the 2006 VHE flares, was triggered by an earlier state of high activity at high energies. Differences between the long-term radio and 
optical light curves might be explained by longer delays in the radio band from synchrotron self absorption and faster cooling times in the optical band. One additional difficulty is that the short-term variations in the optical flux do not seem to be directly connected with variations at high energies. They could arise from the summation of a direct synchrotron component and a smoothed response to the higher energy bands. Optical polarization monitoring would be very useful in probing these different components \citep[e.g.][]{Barr2010}.

\subsection{Hadronic models}

An interesting alternative to leptonic scenarios as considered in this paper is given by hadronic models, where a population of relativistic hadrons, rather than electrons and positrons, is responsible for the dominant contribution to the high-energy emission. However, detailed time-dependent modelling with hadrons is difficult to achieve due to the higher complexity of the hadronic interactions and the large number of free parameters in those models. Moreover, due to the low efficiency of the hadronic emission processes, such scenarios seem generally less adapted to describe the high-energy emission from
blazars \citep{Sik2010}.  

An application of a hadronic scenario to the SED observed during Flare 2 would need to account for the close correlation between the X-ray and VHE flux, the dominance of the VHE emission, the (very) rapid variability time scale, and the low-frequency counterparts. Especially the reproduction of rapid variability presents a problem for hadronic models, since acceleration and cooling time scales for hadrons are much longer than for electrons and positrons. A possible solution to this problem might be given by models where rapid variability is achieved through the interaction of baryonic matter, e.g. in the form of stellar envelopes \citep{Bar2010}, with the blazar jet.

\section{Conclusions}

\label{section:conclusions}

A clear difference in the temporal evolution of the SED has been put in evidence between high- and low-activity states of PKS~2155-304.
The analysis of correlation on various time scales of the flux evolution at different frequencies shows that the steep correlation between X-ray and VHE fluxes already seen during Flare 2 (MJD 53946) is also found in the evolution of the average fluxes over several nights with high VHE flux states. Spectral hardening with an increased flux, already found in the H.E.S.S. data, is also seen in the {\it SWIFT} XRT data, but not in the {\it RXTE} data, owing most likely to the harder energy band observed with {\it RXTE}. 

When the entire available MWL data of the 2006 campaign are considered, there is no compelling evidence of a correlation of the optical flux with the high state in the VHE range on time scales of hours or days. However, the dramatic VHE flaring of PKS~2155-304 in 2006 occurred at a time when the source was already seen in a high-activity state at optical and radio frequencies. Moreover, the optical and radio luminosity further increased after the peak in VHE activity, although on a longer time scale. To the best of our
knowledge, this is the first evidence of a relation between the long-term flux evolution at low frequencies (optical and radio) and VHE events in a blazar. This emphasizes  the present need of global modelling 
of such sources to further explore the still missing link between their low and high-energy behaviour.

The SEDs of the non-flaring nights of the MWL campaign in July 2006 are well reproduced by a stationary one-zone SSC model with only relatively small variations of the parameters that describe the 2003 low state. 
Time dependent multi-zone SSC scenarios arrive at good descriptions of the spectral and flux evolution in the X-ray and VHE bands during Flare 2. They are able to explain the observed dominance of the IC component in the high-energy bands and the steep correlation between VHE and X-ray fluxes with parameters in the generally accepted range for blazars and identify different emission zones. In general, it can be concluded that the SSC framework seems well adapted to describe 
high-energy data from blazars, even in the case of this very rich data set. However, these models still have difficulties to quantitatively reproduce the long-term low-frequency light curves. The more refined models allow in principle for a delayed reaction to the VHE flares in the optical and radio band as observed, but additional emission regions or mechanisms and further modelling of the long-term development of the source would be required to complete the picture and include the whole optical and radio emission in the simulations. Nevertheless, it remains to be seen if yet more refined multi-zone models will ultimately lead to a realistic description of the nature of the emission, or whether the increasing number of free parameters prevent such conclusions, given the finiteness of observable data.

Independent of the model details, however, a distinction of two physically different active VHE states of the blazar PKS~2155-304 is emerging, when considering the present data set together with previously published data on PKS~2155-304. In the low state, generally no correlation is seen between the X-ray and VHE flux, whereas the fluxes observed during high states show a steep correlation. Spectral variability is observed
in the VHE band for low states, but it is different from the spectral hardening seen for high states. The different behaviour during low and high states might be explained within the two-zone scenarios considered in this paper. In this case, the synchrotron emission during low states would be dominated by a not very dense emission region, while a denser and more energetic electron population would account for the emerging correlation and spectral hardening during high states.

\begin{acknowledgements}
The support of the Namibian authorities and of the University of Namibia
in facilitating the construction and operation of H.E.S.S. is gratefully
acknowledged, as is the support by the German Ministry for Education and
Research (BMBF), the Max Planck Society, the French Ministry for Research,
the CNRS-IN2P3 and the Astroparticle Interdisciplinary Programme of the
CNRS, the U.K. Science and Technology Facilities Council (STFC),
the IPNP of the Charles University, the Polish Ministry of Science and 
Higher Education, the South African Department of
Science and Technology and National Research Foundation, and by the
University of Namibia. We appreciate the excellent work of the technical
support staff in Berlin, Durham, Hamburg, Heidelberg, Palaiseau, Paris,
Saclay, and in Namibia in the construction and operation of the
equipment. We also wish to thank L. Foschini for kindly providing the {\it SWIFT} data.
We wish to acknowledge the support of L.~Hanlon, J.~French, and M.~Jelinek 
in obtaining the WATCHER observations. We finally want to thank the anonymous referee
for his helpful comments.
\\
\\
This work is dedicated to the memory of our dear friend and colleague, Okkie de Jager.
\end{acknowledgements}

\clearpage

\appendix
\onecolumn
\section{Tables from the H.E.S.S., {\it RXTE} and {\it SWIFT} XRT spectral analyses}
\label{sec:app0}

   \begin{table*}[h!]
       $$
         \begin{array}{p{0.2\linewidth}l  l  l l}
            \hline
             \noalign{\smallskip}
             T$_0$ &  \phi_0 ( 1 {\mathrm{TeV}})   & \: \: \Gamma & \: \: \chi^2 / \mathrm{ndof} &  \: \: \delta t \\
            \noalign{\smallskip}
            \hline
            \noalign{\smallskip}
        53944.87 &  0.51 \pm 0.06   & \: \: 3.99 \pm 0.12 &  \: \: 3.3 / 6    & \: \: 15.6 \\ 
        53945.87 &  6.62 \pm 0.14   & \: \: 3.45 \pm 0.03 &  \: \: 32.8 / 6  & \: \: 19.0 \\
        53946.88 &   1.90 \pm 0.08  & \: \: 3.77 \pm 0.05  &  \: \: 8.3 / 6  & \: \: 19.0 \\ 
        53948.02 &  0.71 \pm 0.09   & \: \: 4.24 \pm 0.12  &  \: \: 3.3 / 5  & \: \: 8.3 \\
        53949.00  &  1.34 \pm 0.19   & \: \: 3.49 \pm 0.24  &  \: \: 4.5 / 5  & \: \: 7.7 \\  
        53949.99  &  0.73 \pm 0.08   &  \: \: 3.66 \pm 0.13  &  \: \: 3.7 / 6  & \: \: 11.0 \\ 
        53951.03  * &  0.33 \pm 0.06   & \: \: 4.21 \pm  0.19 &  \: \: 2.6 / 5   & \: \: 14.2 \\
            \noalign{\smallskip}
            \hline
            quiescent & 0.18 \pm 0.13 & \: \: 3.53 \pm 0.06 &  \: \:  & \: \: - \\
            \hline
         \end{array} 
      $$
        \caption[]{Average H.E.S.S. spectral parameters during the nights where X-ray data are available. The start time of observations T$_0$ (in MJD), the flux normalization 
        $\phi_0$ at 1 TeV (in 10$^{-11}$  cm$^{-2}$ s$^{-1}$ TeV$^{-1}$), the photon index $\Gamma$ for a power law, the $\chi^2$ over numbers of degree of freedom (ndof)
         and the live time of the selected runs (in ks) are listed. * The entry with the start time 53951.03 is an average over three nights (MJD 53951 to 53953). The quiescent flux
         is taken from~\cite{vhe_paper}. }
        \label{tab:vhe}
  \end{table*}

   \begin{table*}[h!]
       $$
         \begin{array}{p{0.2\linewidth}l  l  l l }
            \hline
             \noalign{\smallskip}
              T$_0$  &  \phi_0 (1 \mathrm{keV})   & \: \:  \Gamma & \: \: \chi^2 / \mathrm{ndof} & \: \: \delta t \\
            \noalign{\smallskip}
            \hline
            \noalign{\smallskip}              
            53945.04  & 0.05 \pm  0.01 & \: \: 3.06 \pm  0.18 & \: \: 12.6 / 24  & \: \: 0.7 \\ 
            53946.02  & 0.16 \pm 0.01 &  \: \: 2.97 \pm 0.06 & \: \: 22.1 / 24  & \: \: 0.8 \\
            53946.08  & 0.15 \pm 0.02 &  \: \: 3.02 \pm  0.08 & \: \: 16.4 / 24  &\: \: 0.5  \\
            53947.00 & 0.13 \pm 0.02 & \: \: 3.08 \pm 0.08 & \: \: 14.6 / 24  & \: \: 0.9 \\
            53947.07  & 0.19 \pm 0.03 & \: \: 3.28 \pm 0.09 & \: \: \:\: 8.7 / 24  &\: \: 0.6 \\ 
            53948.05  & 0.10 \pm 0.01 & \: \: 2.95 \pm 0.09 & \: \: 22.6 / 24 &\: \: 0.7 \\
            53949.03  & 0.09 \pm 0.01 &  \: \: 3.03 \pm 0.11 & \: \:  19.1 / 24 &\: \: 0.8 \\                                                                                               
            53950.08  & 0.12 \pm  0.02 &  \: \: 3.02 \pm 0.09 & \: \:  28.3 / 24  &\: \: 0.6 \\
            53951.06  & 0.09 \pm  0.02 & \: \: 3.27 \pm 0.16 & \: \:  18.4 / 24  &\: \: 0.7 \\
            \hline
         \end{array} 
      $$
        \caption[]{{\it RXTE} spectral parameters from the different pointings of the nights during the H.E.S.S. campaign. The start time of observations T$_0$ (in MJD), the extrapolated 
        flux normalization $\phi_0$ at 1 keV (in cm$^{-2}$ s$^{-1}$ keV$^{-1}$), the photon index $\Gamma$ for a power law, the $\chi^2$ and ndof of the fit,
         and the live time of each pointing (in ks) are listed.}
      \label{tab:rxte}
  \end{table*}    

  \begin{table*}[h!]
       $$
       \begin{array}{p{0.14\linewidth}l l l l l l}
            \hline
             \noalign{\smallskip}
               T$_0$  &  \phi_0 (1 \mathrm{keV})  & \:  \Gamma_{\mathrm{soft}}  & \:\: E_{\mathrm{break}}  & \: \Gamma_{\mathrm{hard}}  & \: \: \chi^2 / \mathrm{ndof} & \: \delta t \\
            \noalign{\smallskip}
            \hline
            \noalign{\smallskip}              
            53945.04 &  0.02 \pm 0.001  &  \:  2.84  \pm  0.09 & - & -  & 201.3 / 185   & \: 0.8 \\ \noalign{\smallskip}
                              & 0.02 \pm 0.002    & \; 2.72 \pm 0.17     & 1.0 & 2.96 \pm 0.19 & 200.8 / 184 & \; 0.8 \\ \noalign{\smallskip}
            53946.04 &  0.08 \pm  0.001 &  \:  2.40  \pm  0.02 & 1.00 \pm 0.06 & \: 2.70 \pm 0.02  & 371.0 / 365 & \: 4.8 \\ \noalign{\smallskip}
            53948.05 &   0.06 \pm  0.003 & \:  2.59 \pm 0.07  &  1.04 \pm 0.32  & \:    2.84 \pm 0.10  & 87.4 / 90 & \:  0.4 \\     \noalign{\smallskip}
            53949.00 &    0.05 \pm  0.001 &  \:  2.58 \pm  0.03 & 1.18 \pm 0.15  & \:  2.84 \pm 0.05  & 180.1 / 193 & \: 1.8 \\ \noalign{\smallskip}
            53950.06 &     0.06 \pm  0.001 &  \: 2.37 \pm  0.04 & 1.01 \pm 0.10 & \:   2.70 \pm 0.04  & 242.4 / 202 & \: 1.6 \\  \noalign{\smallskip}
            53952.13 &     0.04 \pm  0.001 &  \: 2.73 \pm  0.03 &  -  & \:             -               &  77.0 / 94 & \: 0.5 \\    \noalign{\smallskip}
                               &    0.04 \pm 0.001 & \; 2.62 \pm 0.06 & 1.0  & 2.91 \pm 0.08       &  71.0 / 93 & \: 0.5 \\    \noalign{\smallskip}
            53953.15 &     0.03 \pm 0.001 &  \: 2.74 \pm 0.05  &  -  & \:              -               &  47.1 / 52 &\: 0.3 \\   \noalign{\smallskip}
            \hline
         \end{array} 
      $$
        \caption[]{{\it SWIFT}-XRT spectra of the nights during and after the H.E.S.S. campaign. The start time of observations T$_0$ (in MJD), the 
        flux normalization $\phi_0$ at 1 keV (in cm$^{-2}$ s$^{-1}$ keV$^{-1}$), the photon indices below ($\Gamma_{\mathrm soft}$) and above ($\Gamma_{\mathrm hard}$) a spectral break (E$_{\mathrm   break}$ in keV),  the $\chi^2$ over numbers of degree of freedom (ndof) and the live time of each pointing (in ks) are listed. Where no break energy is indicated, the third column provides the photon index over the whole spectrum. For the pointings on MJD 53945 and MJD 53952, 
         the results of fits with a power law and with a broken power law with fixed break energy are shown. For the pointing on MJD 53953, a fit with a broken power law did not yield a
         satisfactory result, even when fixing the break energy. } 
      \label{tab:xrt}
  \end{table*}    

\clearpage

\twocolumn
\section{A short  description of the time-dependent models}

\subsection{Model 1}
\label{subsec:app1}

In Model 1, a power-law distribution of relativistic electrons with index 2, a natural outcome of first order Fermi acceleration, is injected into the source volume, divided into several homogeneous cubic cells. The time development of these cells is simulated, including electron injection, radiative cooling, synchrotron and SSC emission, as well as absorption of light traversing the cells due to electron-positron pair production. Adiabatic expansion of the emission
region is not taken into account.

The model does account for the light crossing time effect between the cells and a comoving observer, which becomes significant in the case of rapid variability time scales. The comoving observer is assumed to view the jet under an angle of 90$^\circ$, corresponding to a maximum delay of the light travel time for cells at different locations in the jet. The observed synchrotron spectrum is a superposition of spectra from different parts of the emission region and thus from electron distributions that have had more or less time to cool. This superposition leads to a break in the observed spectrum in a natural way, contrary to the stationary model discussed in Section~\ref{subsec:stationary}, where a break is introduced {\it a priori} in the electron energy distribution.

A more detailed description of the model is given by \citet{Chi1999} and \citet{Kat2008}. Table~\ref{tab:dynamic1} shows the set of parameters that were used for a description of the varying SED during Flare 2.

\begin{table} [h]
  \centering
    \begin{tabular}{p{0.5\linewidth} c}
     \hline  
     \multicolumn{2}{ c }{\mbox{source 1 (dominant in X-rays)}} \\
    \hline 
    $\delta_{1}$ & 20 \\
    $ B_{1} [\rm{G}] $ & $ 0.06 $ \\
    $ L_{1} [\rm{cm}] $ & $ 4.0 \times 10^{16} $ \\
    $ K_{1} [\rm{cm}^{-3}] $ & $1.6 \times 10^{4} $\\
    $ n_{1} $ & 2.0  \\
    $  \gamma_{1}^{\rm min} $ & 1.0  \\
    $  \gamma_{1}^{\rm max} $ & $6.0 \times 10^{5}$ \\
    \hline
     \multicolumn{2}{ c }{\mbox{source 2 (dominant in VHE)}} \\
     \hline
    $\delta_{2}$ & 30 \\
    $ B_{2} [\rm{G}] $ & 0.05 \\
    $  L_{2} [\rm{cm}] $& $2.35 \times 10^{15}$ \\
    $ K_{2} [\rm{cm}^{-3}] $ & $1.6 \times 10^{6} $\\
    $ n_{2} $ & 2.0 \\
    $  \gamma_{2}^{\rm min} $ & 1.0  \\
    $  \gamma_{2}^{\rm max} $ & $1.0 \times 10^{6}$ \\
    \hline
  \end{tabular}
  \caption{Principal parameters used in Model 1 for a description of Flare 2. A definition of the parameters can be found in~\cite{Kat2008}. }
  \label{tab:dynamic1}
\end{table}

\subsection{Model 2}
\label{subsec:app2}

In this model, the high-energy emission comes from relativistic electrons and positrons in an extended, inhomogeneous jet and in a dense plasma blob travelling along the jet axis.  

The jet is at the origin of the long-term component, modelled as a steady radiation, that dominates the X-ray synchrotron emission. It has a global paraboloidal shape and is approximated by a continuous outflow of homogeneous cylindrical slices from the central engine, which travel with constant velocity and expand adiabatically. The magnetic field intensity decreases along the jet. Adiabatic and synchrotron cooling and particle escape are accounted for, whereas the weak radiation field inside the extended jet makes IC cooling negligible. 

The inner plasma blob is a denser and more energetic zone, which is at the origin of the VHE flux and of the rapid variability through its SSC emission. The blob is built of homogeneous slices, in the same way as the jet. In the blob, 
the higher radiation field energy density implies a significant IC cooling, in addition to the cooling processes already considered for the particles in the jet. 

Radiative transfer along the jet is treated under the assumption of a small viewing angle in the comoving frame, thus the light crossing time effect described in Model 1 is negligible here. In addition to the radiation transfer inside each emission region, Model 2 also takes into account the absorption of synchrotron radiation from the blob by jet particles and vice versa, as well as the absorption of IC photons emitted from the blob through electron-positron pair production in the jet. A particle distribution is injected at the base of the jet following a power-law with initial spectral index $n_{\rm jet}$ in the jet and $n_{\rm blob}$ in the blob. 

Synchrotron emission by an additional component, identified as a large kpc-scale jet, dominates the radio and optical emission and completes the picture of the observed MWL emission from the blazar. Emission from this large-scale jet is treated here as independent from the other components, i.e. the inner jet and plasma blob.

A more detailed description of this model is given by \citet{Kat2003} and \citet{Len2009a}. Table~\ref{tab:dynamic2} shows the set of parameters that were used for a description of the varying SED during Flare 2.

\begin{table}[h!]
  \centering
    \begin{tabular}{p{0.5\linewidth} c}
     \hline  
     \multicolumn{2}{ c }{\mbox{plasma blob}} \\
     \hline
    $\delta_{\rm blob}$ & 50 \\
    $ B_{\rm blob(1)} [{\rm G}] $ & $ 0.03 $ \\
    $ R_{\rm blob(1)} [{\rm cm}] $ & $ 6.0 \times 10^{15} $ \\
    $ K_{\rm blob(1)} [{\rm cm}] $ & $ 9.7 \times 10^{6} $\\ 
    $  \gamma_\mathrm{blob}^{\rm min} $& $1.0 \times 10^{3}$  \\
    $  \gamma_{\rm blob}^{\rm max} $ &  $7.4 \times 10^{5}$ \\
   $ n_{\rm blob} $ & 2.5 \\
    \hline
         \multicolumn{2}{ c }{\mbox{inner jet}} \\
    \hline 
    $\delta_{\rm jet}$ & 50 \\
    $ B_{\rm jet(1)} [\rm{G}] $ & $ 0.01  $ \\
    $ R_{\rm jet(1)} [{\rm cm}] $ & $ 6.0 \times 10^{16} $ \\
    $ K_{\rm jet(1)} [{\rm cm}^{-3}] $ & $2.3 \times 10^{3} $\\
    $  \gamma_\mathrm{jet}^{\rm min} $& 1.0  \\
    $  \gamma_{\rm jet}^{\rm max} $ & $4.0 \times 10^{5}$ \\
    $ n_{\rm jet}$ & 2.0 \\
    \hline
  \end{tabular}
  \caption{Principal parameters used in Model 2 for a description of the high-energy emission of Flare 2. A detailed description of the parameters can be found in~\citet{Kat2003}.
   The parameters used for the large-scale jet that dominates the radio and optical emission are the same as given in Table~\ref{tab:stationary} in Section~\ref{subsec:stationary}.}
  \label{tab:dynamic2}
\end{table}

\subsection{Model 3}
\label{subsec:app3}

The third model fits well into the two-flow framework originally proposed by \citet{Pel85} and \citet{Sol89} (see also \citealt{Tsin02} or the "spine-in-jet" model developped by \citealt{Chia00}) where a non relativistic but powerful MHD jet launched by the accretion disk surrounds a highly relativistic plasma of electron-positron pairs propagating along its axis. The MHD jet plays the role of a collimator and energy reservoir for the pair plasma, which is responsible for the observed broad-band emission.

The jet has a paraboloidal shape and its magnetic field decreases along the axis following a power law. A continuous acceleration of the plasma is assumed, leading to an increase of the bulk Doppler factor up to an asymptotic value 
$\delta_{\rm b\infty}$, over an acceleration region starting at the base of the jet. The pair plasma is continuously being injected at the base of the jet, and its evolution is computed as it propagates inside the MHD structure.  The plasma is re-accelerated along the jet to compensate for synchrotron and IC cooling, which are responsible for the observed high-energy emission, following an SSC scenario. Particle acceleration is supposed to be provided by the interaction of the plasma with the magnetic turbulence carried by the surrounding MHD structure and is described as a power law of index $\zeta$: $Q_{\rm{acc}}(z)\propto Q_{0}(t)\,(z/Z_{0})^{-\zeta}$. Hence, the particle energy distribution is a relativistic Maxwellian  (or ``pile-up'') distribution, which is the natural outcome of second order Fermi acceleration. This is expected for example in the case of particle-wave interaction or reconnection (\citealt{Sch84,Sch85}; \citealt{hen91}; \citealt{SH04}; \citealt{Bou2008}). 

The evolution of the observed flux is reproduced as a consequence of time dependent injection parameters. Hence, by varying the acceleration term at the base of the jet and the flux of injected particles, a time dependent SED
and light curves in the different energy bands can be reproduced. Given the narrow energy distribution and the effects of cooling of the plasma along the jet, the resulting SED is a superposition of contributions from different regions in the jet, where emissivity peaks at lower energies for particles that have travelled farther.
The high-energy peaks are dominated by emission close to the base of the jet.

Contrary to the previously described two models, pair production is not only calculated to account for losses of VHE $\gamma$-rays, but the first generation of electron-positron pairs is added to the 
particle flux and re-accelerated with the injected plasma. This leads to an enhancement of initial variations of the particle density, in particular during a flare. 

A more detailed description of this model is given by ~\citet{Bou2008} and~\citet{Bou2009}. Table~\ref{tab:dynamic3} shows the set of parameters that were used for a description of the varying SED during Flare 2.

\begin{table}[h] 
  \centering
    \begin{tabular}{p{0.3\linewidth}p{0.1\linewidth}lc}
     \hline  
     \multicolumn{4}{ c }{\mbox{jet parameters}} \\
    \hline 
    $ \delta_{\rm b\infty}$ & & & 30\\
    $ R_{\rm i}\, [{\rm cm}] $ &  &  & $ 1.1\times10^{14} $ \\
    $ R_{0}\, [{\rm cm}] $ &  &  & $ 1.78 \times 10^{14} $ \\
    $ Z_{0} \,[{\rm cm}] $ &  &  & $12 \times 10^{15} $\\
    $ Z_{\rm c} \,[{\rm cm}] $ &  &  & $ 2 \times 10^{20} $\\
    $ B_{0}\, [{\rm G}] $ &  &  & $ 4.5 $\\
    $ \omega $ &  &  & $ 0.16 $\\
    $ \lambda $ &  &  & $ 1.8 $\\
    $ \zeta $ &  &  & $ 1.18 $\\
    \hline
  \end{tabular}
\caption{Principal parameters used in Model 3 for the description of Flare 2. A definition of these parameters and details on constraints from observation can be found in 
\citet{Bou2008} and \citet{Bou2009}.}
\label{tab:dynamic3}
\end{table}

\end{document}